\documentclass[iop,apjl,revtex4,tighten]{emulateapj}
\usepackage{natbib}
\usepackage{times}
\usepackage{color}
\usepackage{gensymb}

\newcommand{\msun}{\ensuremath{M_\sun}}
\newcommand{\msuns}{\ensuremath{M_\sun\,{\rm s}^{-1}}}

\newcommand{\tadv}{\ensuremath{\tau_{\rm adv}}}
\newcommand{\theat}{\ensuremath{\tau_{\rm heat}}}

\newcommand{\nue}{\ensuremath{\nu_{e}}}
\newcommand{\nuebar}{\ensuremath{\bar \nu_e}}
\newcommand{\numt}{\ensuremath{\nu_{\mu\tau}}}
\newcommand{\numtbar}{\ensuremath{\bar \nu_{\mu\tau}}}
\newcommand{\numu}{\ensuremath{\nu_{\mu}}}
\newcommand{\nutau}{\ensuremath{\nu_{\tau}}}
\newcommand{\numubar}{\ensuremath{\bar \nu_{\mu}}}
\newcommand{\nutaubar}{\ensuremath{\bar \nu_{\tau}}}
\newcommand{\mev}{\mbox{MeV}}

\newcommand{\Rshock}{\ensuremath{R_{\rm shock}}}

\newcommand{\Ediag}{\ensuremath{E^{+}}}

\newcommand{\alp}{\ensuremath{\alpha}}

\newcommand{\kbbar}{\ensuremath{\rm k_B \; baryon^{-1}}}

\newcommand{\heateff}{\ensuremath{\eta_{\rm heat}}}
\newcommand{\accgain}{\ensuremath{\dot{M}_{\rm gain}}}
\newcommand{\accshock}{\ensuremath{\dot{M}_{\rm sh}}}

\newcommand{\gcc}{\ensuremath{{\mbox{g~cm}}^{-3}}}

\newcommand{\kmps}{\ensuremath{\mbox{km~s}^{-1}}}

\newcommand{\Bethes}{\ensuremath{{\mbox{B~s}}^{-1}}}

\newcommand{\nurhd}{{$\nu$RHD}}
\newcommand{\ndc}{{$\nu$DC}}

\newcommand{\chimera}{{\sc Chimera}}

\newcommand{\Eturb}[1]{\ensuremath{E_{\rm turb}^{\rm #1}}}
\newcommand{\vturb}[1]{\ensuremath{\mathbf{v}^{\rm #1}}}

\slugcomment{Accepted for publication in ApJ Letters: June 9, 2015}

\shorttitle{3D core-collapse supernova simulation}
\shortauthors{Lentz et al.}

\begin{document}

\title{Three-dimensional core-collapse supernova simulated using a 15~\msun\ progenitor}
\author{
Eric J. Lentz\altaffilmark{1,2},
Stephen W. Bruenn\altaffilmark{3},
W. Raphael Hix\altaffilmark{2,1},
Anthony Mezzacappa\altaffilmark{1,4},
O. E. Bronson Messer\altaffilmark{5,2,1},\\
Eirik Endeve\altaffilmark{6,1,4},
John M. Blondin\altaffilmark{7},
J. Austin Harris\altaffilmark{2},
Pedro Marronetti\altaffilmark{8}, and
Konstantin N. Yakunin\altaffilmark{1,2,4}
}
\email{elentz@utk.edu}

\altaffiltext{1}{Department of Physics and Astronomy, University of Tennessee, Knoxville, TN 37996-1200, USA}
\altaffiltext{2}{Physics Division, Oak Ridge National Laboratory, P.O. Box 2008, Oak Ridge, TN 37831-6354, USA}
\altaffiltext{3}{Department of Physics, Florida Atlantic University, 777 Glades Road, Boca Raton, FL 33431-0991, USA}
\altaffiltext{4}{Joint Institute for Computational Sciences, Oak Ridge National Laboratory, P.O. Box 2008, Oak Ridge, TN 37831-6173, USA}
\altaffiltext{5}{National Center for Computational Sciences, Oak Ridge National Laboratory, P.O. Box 2008, Oak Ridge, TN 37831-6164, USA}
\altaffiltext{6}{Computer Science and Mathematics Division, Oak Ridge National Laboratory, P.O. Box 2008, Oak Ridge, TN 37831-6164, USA}
\altaffiltext{7}{Department of Physics, North Carolina State University,  Raleigh, NC 27695-8202, USA}
\altaffiltext{8}{Physics Division, National Science Foundation, Arlington, VA 22207 USA}

\begin{abstract}
We have performed \emph{ab initio} neutrino radiation hydrodynamics simulations in three and two spatial dimensions (3D and 2D) of core-collapse supernovae from the same 15~\msun\ progenitor through 440~ms after core bounce.
Both 3D and 2D models achieve explosions, however, the onset of explosion (shock revival) is delayed by $\sim$100~ms in 3D relative to the 2D counterpart and the growth of the diagnostic explosion energy is slower.
This is consistent with previously reported 3D simulations utilizing iron-core progenitors with dense mantles.
In the $\sim$100~ms before the onset of explosion, diagnostics of neutrino heating and turbulent kinetic energy favor earlier explosion in 2D.
During the delay, the angular scale of convective plumes reaching the shock surface grows and explosion in 3D is ultimately lead by a single, large-angle plume, giving the expanding shock a directional orientation not dissimilar from those imposed by axial symmetry in 2D simulations.
We posit that shock revival and explosion in the 3D simulation may be delayed until sufficiently large plumes form, whereas such plumes form more rapidly in 2D, permitting earlier explosions.
\end{abstract}


\section{Introduction}

That massive stars explode at the end of their lives is well established observationally \citep{Smar09}.
Numerical simulation of core-collapse supernovae (CCSNe) has been less consistently successful than Nature \citep[e.g.,][]{Jank12,HiLeEn14,MeBrLe15}.
The ultimate source of the neutrino-driven explosion mechanism is the conversion of the gravitational binding energy of the core, collapsed to a proto-neutron star (proto-NS), and of matter accreted onto the proto-NS, into neutrinos that heat material behind the shock --- reviving it and expelling the stellar envelope as a supernova.
This process is decidedly non-spherical.
Neutrino heating above the proto-NS drives neutrino-driven convection \citep[\ndc;][]{Beth90,HeBeCo92,HeBeHi94}.
Also excited are the low-order modes of the standing accretion shock instability \citep[SASI;][]{BlMeDe03}.
Asphericities in the shock surface channel the continuing accretion into distinct streams.
All of these emergent behaviors are manifestly different with imposed axisymmetry (2D) than without (3D); therefore, we should expect 3D modeling to impact the initiation and subsequent development of CCSNe.

Fully capturing the complex behaviors of the CCSN central engine numerically requires spectral neutrino transport, as the neutrinos are not in equilibrium with the fluid and heating is neutrino energy dependent, coupled to self-gravitating hydrodynamics; i.e., spectral neutrino radiation hydrodynamics (\nurhd).
Few self-consistent, spatially 3D, spectral-\nurhd\ CCSN simulations have been reported.
\citet{TaKoSu12} computed a low-resolution 3D simulation of an 11.2-\msun\ progenitor through the start of explosion.
A 2.7$\degree$ angular resolution follow-up using the same progenitor \citep{TaKoSu14} also initiated explosions that subsequently developed more slowly than corresponding 2D simulations.
\citet{HaMuWo13} computed a non-exploding 2.0$\degree$ 3D simulation of a 27-\msun\ progenitor, though the same progenitor and physics did explode in 2D.
3D simulations of 11.2- and 20-\msun\ progenitors with the same code and configuration also failed to explode \citep{TaHaJa14}, though \citet{MeJaBo15} found a delayed 3D explosion for a 20-\msun\ model with modified opacities.
This pattern of delayed and failed explosions in 3D was suggested by \citet{HaMaMu12} and later demonstrated explicitly \citep{Couc13b,CoOc14} using simulations with parameterized treatments of neutrino heating and cooling, though other simulations showed only small differences or the opposite pattern \citep{NoBuAl10,DoBuMu13,HaPlOd14,Fern15}.
Earlier work by \citet{FrWa02,FrWa04} using non-spectral \nurhd\ found explosions in 2D and 3D to be similar.

In this \emph{Letter} we report on the first 440~ms of post-bounce evolution for a CCSN initiated from a 15-\msun\ progenitor in 3D and 2D using our multi-dimensional, \nurhd\ code \chimera\footnote{\url{ChimeraSN.org}} (Bruenn et al., in prep.) with modern neutrino interactions and general relativistic corrections to Newtonian self-gravity.
We find that the shock revival occurs earlier by $\sim$100~ms in the 2D simulation relative to 3D, and that the growth of the diagnostic explosion energy is similarly accelerated, potentially resulting in stronger explosions in 2D than 3D.
This is the first reported 3D explosion at 15~\msun, a representative mass often used for comparative studies, and the first for \chimera.
In Section~\ref{sec:methods}, we summarize our methodology and initial conditions.
An overview of the simulations is presented in Section~\ref{sec:overview} with a focus on the differences between the 2D and 3D simulations in Section~\ref{sec:2d3d}.
We discuss our results in context in Section~\ref{sec:discuss} followed by a summary in Section~\ref{sec:summary}.

\section{Numerical methods and inputs}
\label{sec:methods}

\defcitealias{BrMeHi13}{B2013}
\defcitealias{BrLeHi14}{B2014}

Initial conditions are taken from the 15~\msun\ pre-supernova progenitor of \citet{WoHe07}.
The inner region (10,700 km; 2.32~\msun) is remapped onto 540 radial shells on logarithmic radial grid ($\delta r/r$) modified to track density gradients.
Multi-dimensional simulations were initialized from a 1D simulation at 1.3~ms after bounce by applying a 0.1\% random density perturbation over radii 10--30~km, mimicking perturbations seen in simulations evolved through bounce in 2D.
The angular grid of the 3D simulation (C15-3D) was initialized with a 180-zone ($\Delta\phi=2\degree$) $\phi$-grid and a 180-zone $\theta$-grid equally spaced in $\mu \equiv \cos\theta$, i.e., equal solid angle.
This $\theta$-grid widens the pole-adjacent zones ($\Delta \ell = R_{\rm sph} \Delta\phi \sin \theta $) and therefore the time step .
We evolve in spherical symmetry inside $R_{\rm sph} = 6$~km until 45~ms after bounce (when prompt convection fades) thereafter setting $R_{\rm sph} = 8$~km.
With this grid, the pole-most zone is $\approx$8.5\degree\ wide resulting in a minimum length and time step $\approx$4$\times$ larger than for a uniform $2\degree$ $\theta$-grid \citep[e.g.,][]{HaMuWo13}.
300~ms after bounce, the $\theta$-grid was remapped in the 10 $\theta$-zones closest to each pole ($\approx$27\degree) to uniform spacing  ($\Delta\theta=2.7\degree$) and the $\phi$-sweep at the pole was replaced by averaging, yielding similar time steps.
The axisymmetric simulation (C15-2D) uses 270 uniform $\theta$-zones ($\Delta\theta=2/3\degree$).

\begin{figure}
\includegraphics[width=\columnwidth,clip]{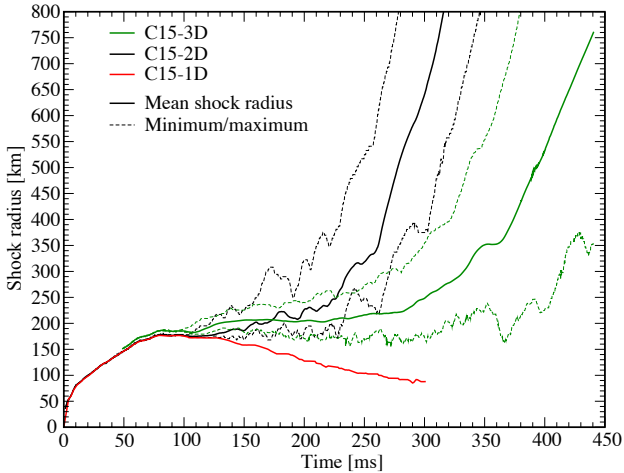}
\caption{\label{fig:shockradii}
Mean (solid) shock radius for models C15-3D (green), C15-2D (black), and C15-1D (red) plotted versus time. Minima and maxima plotted with dashed lines.
}
\end{figure}

These are the third series of \chimera\ simulations (Series-C) and are substantially similar to the Series-B simulations \citep[][hereafter \citetalias{BrMeHi13} and \citetalias{BrLeHi14}]{BrMeHi13,BrLeHi14}.
A more extensive description of \chimera\ can be found in \citet{BrLeHi14}.
The included microphysics are the same as for the Series-B models including the spherical GR terms in the gravity and transport.
We solve the multi-group flux-limited diffusion equations for all three flavors of neutrinos and anti-neutrinos with four coupled species: \nue, \nuebar, $\numt=\{\numu,\nutau\}$, $\numtbar=\{\numubar,\nutaubar\}$, using 20 logarithmically spaced energy groups $\alpha\epsilon = 4$--250~\mev, where $\alpha$ is the lapse function and $\epsilon$  the comoving-frame group-center energy, in the ray-by-ray approximation.
The neutrino--matter interactions used are the full set of \citetalias{BrLeHi14}.
We utilize the \citet{LaSw91} EoS (incompressibility $K = 220$~\mev) for  $\rho>10^{11}$~\gcc\ and an enhanced version of the \cite{Coop85} EoS for  $\rho<10^{11}$~\gcc, and in outer regions a 14-species \alp-network \citep{HiTh99b}.

Relative to the Series-B simulations \citepalias{BrMeHi13,BrLeHi14}, the neutrino transport solver now corrects for frame differences between shock-adjacent zones when computing the flux and flux gradients (S.~W. Bruenn et al., in prep.), permitting spherically symmetric \chimera\ simulations to track the late shock retreat of the reference simulation in \citet{LeMeMe12b}.
This improvement has a modest effect on the shock stalling radius.

All times are given relative to core bounce.
The proto-NS is defined as the volume where $\rho>10^{11}$~\gcc\ and the shocked `cavity' is the volume between the proto-NS and the shock.

\section{Simulation overview}
\label{sec:overview}

After remapping from 1D, the multi-D simulations proceed in similar fashion: convectively unstable regions left behind by the shock progress through the  Fe-core trigger prompt convection inside the proto-NS, similar to the axisymmetric Series-B simulations.

\begin{figure}
\includegraphics[width=\columnwidth,clip]{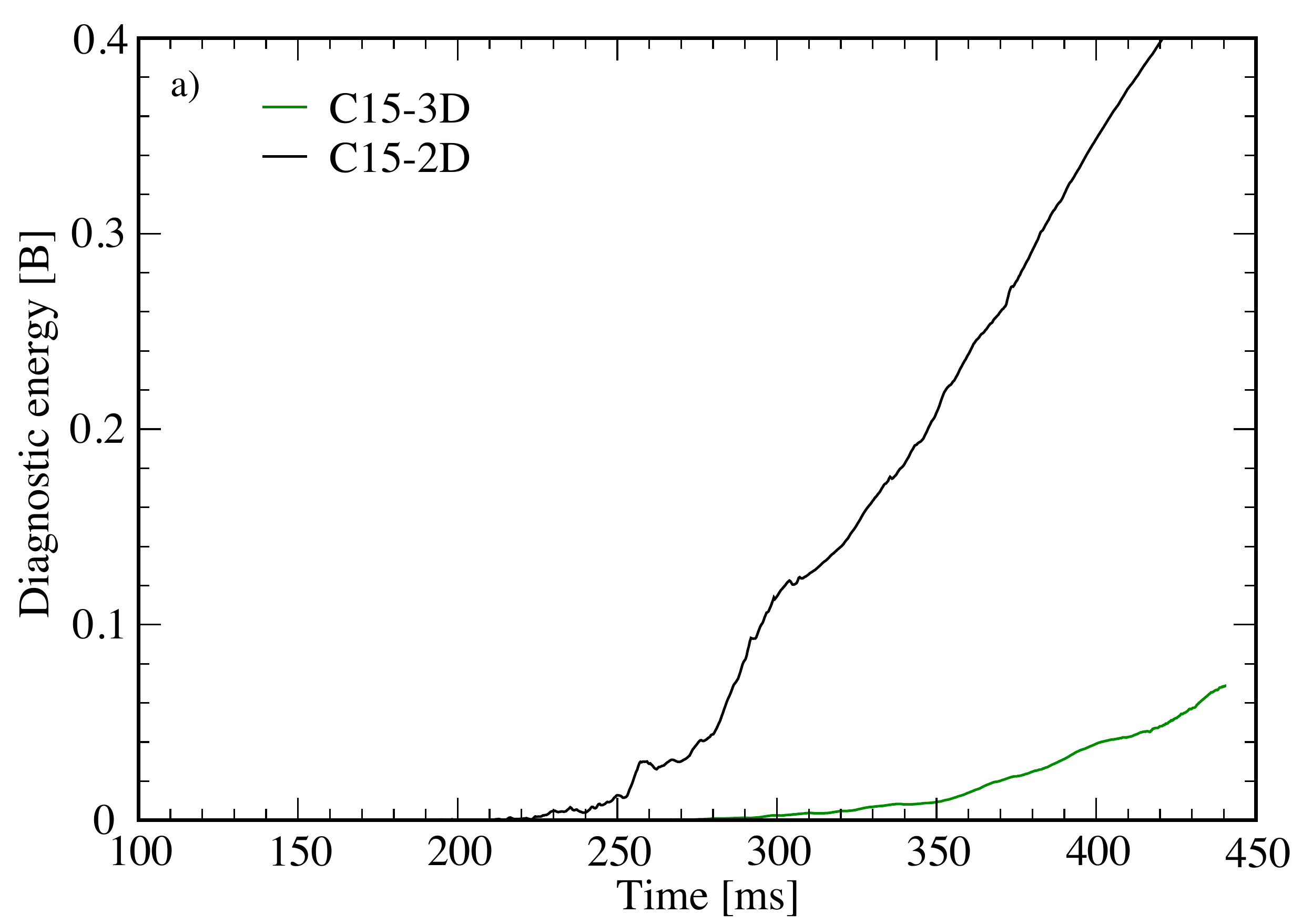}
\includegraphics[width=\columnwidth,clip]{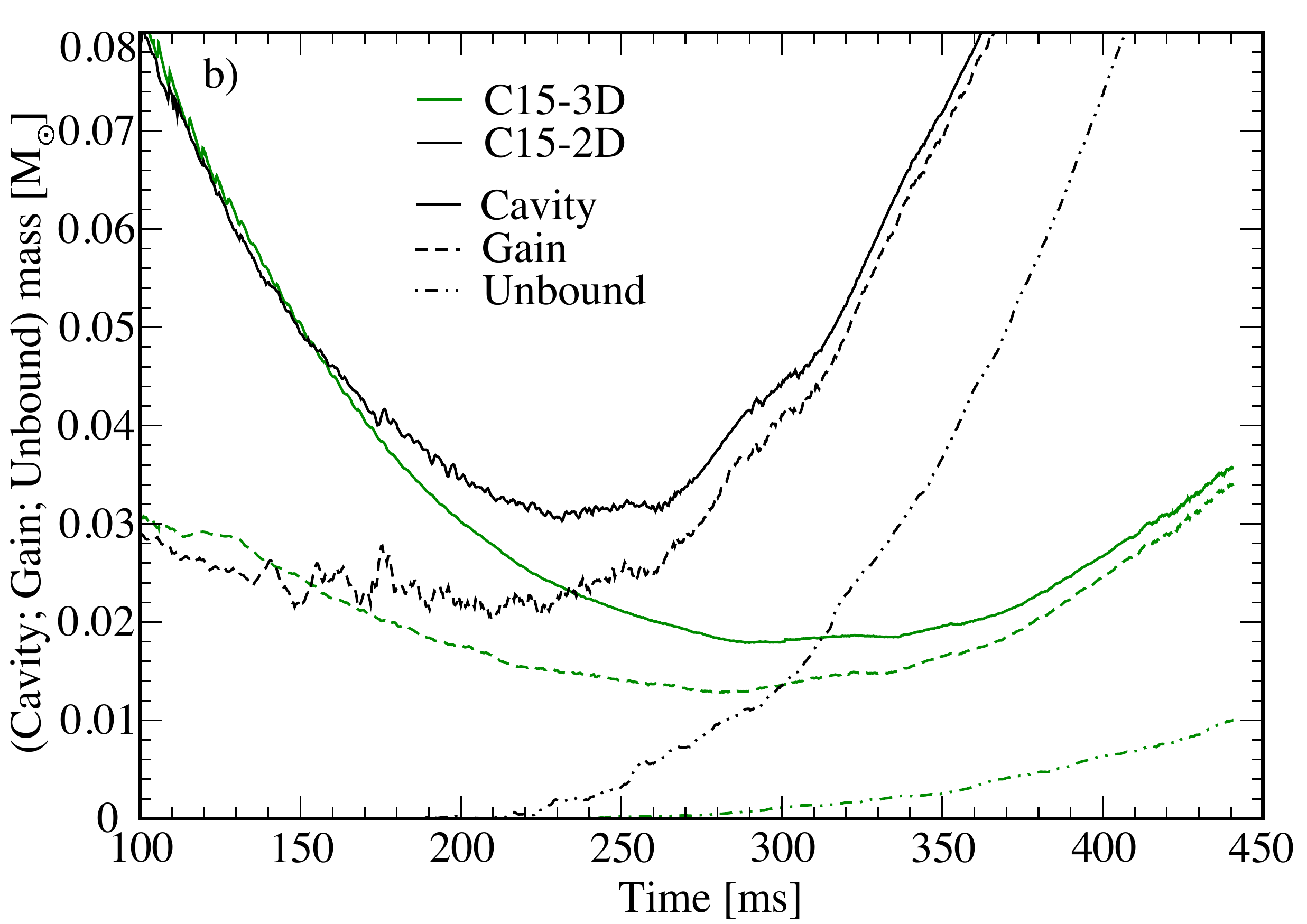}
\caption{\label{fig:overview}
Panel a: Diagnostic energy \Ediag, and
Panel b: Mass of shocked cavity (solid), gain region (dashed), and unbound region (dash-dotted) plotted in colors of Figure~\ref{fig:shockradii}. See text and \citetalias{BrLeHi14} for definitions.
}
\end{figure}

\begin{figure*}
\begin{center}
\includegraphics[width=.95\columnwidth]{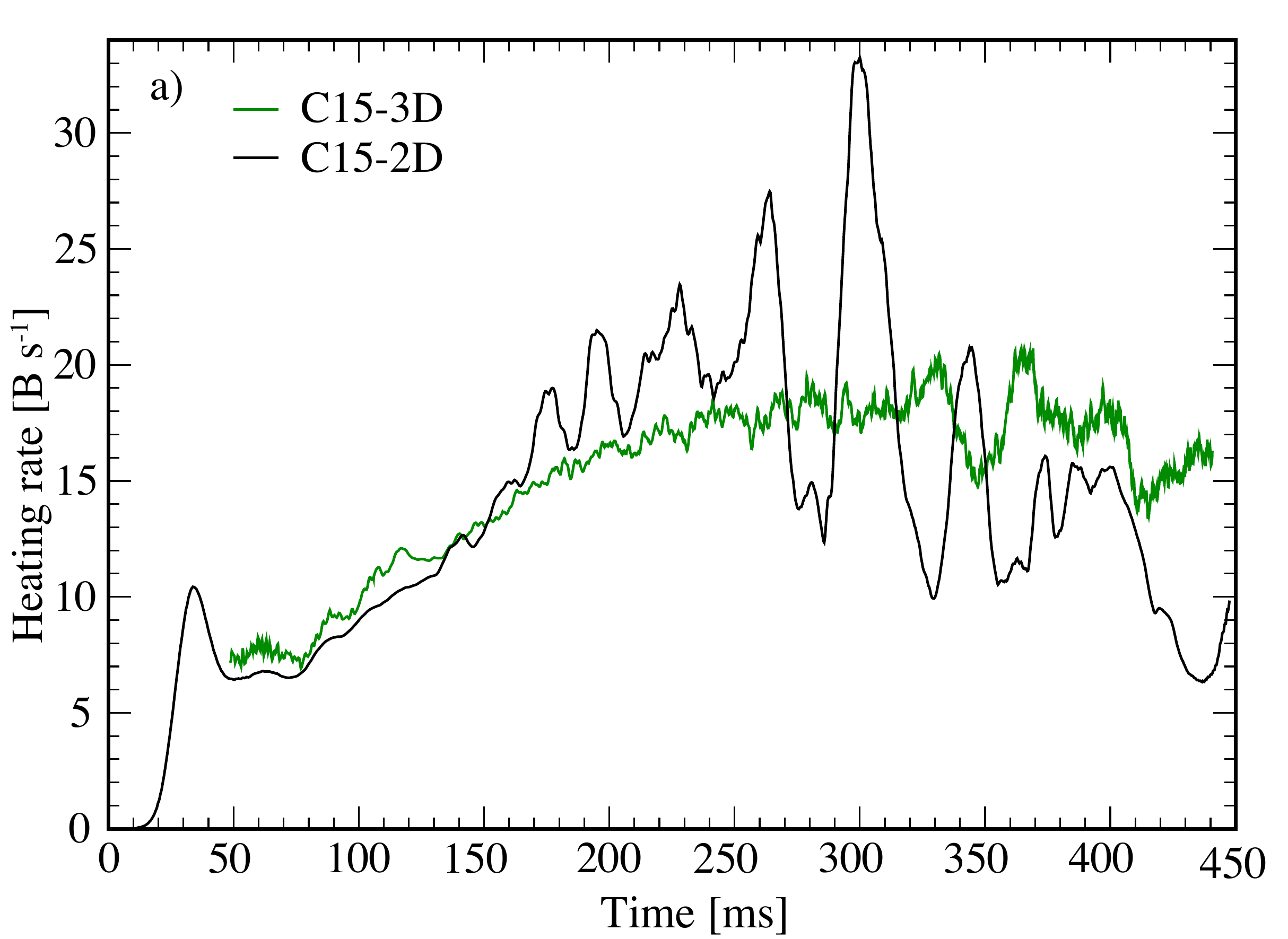}
\includegraphics[width=.95\columnwidth]{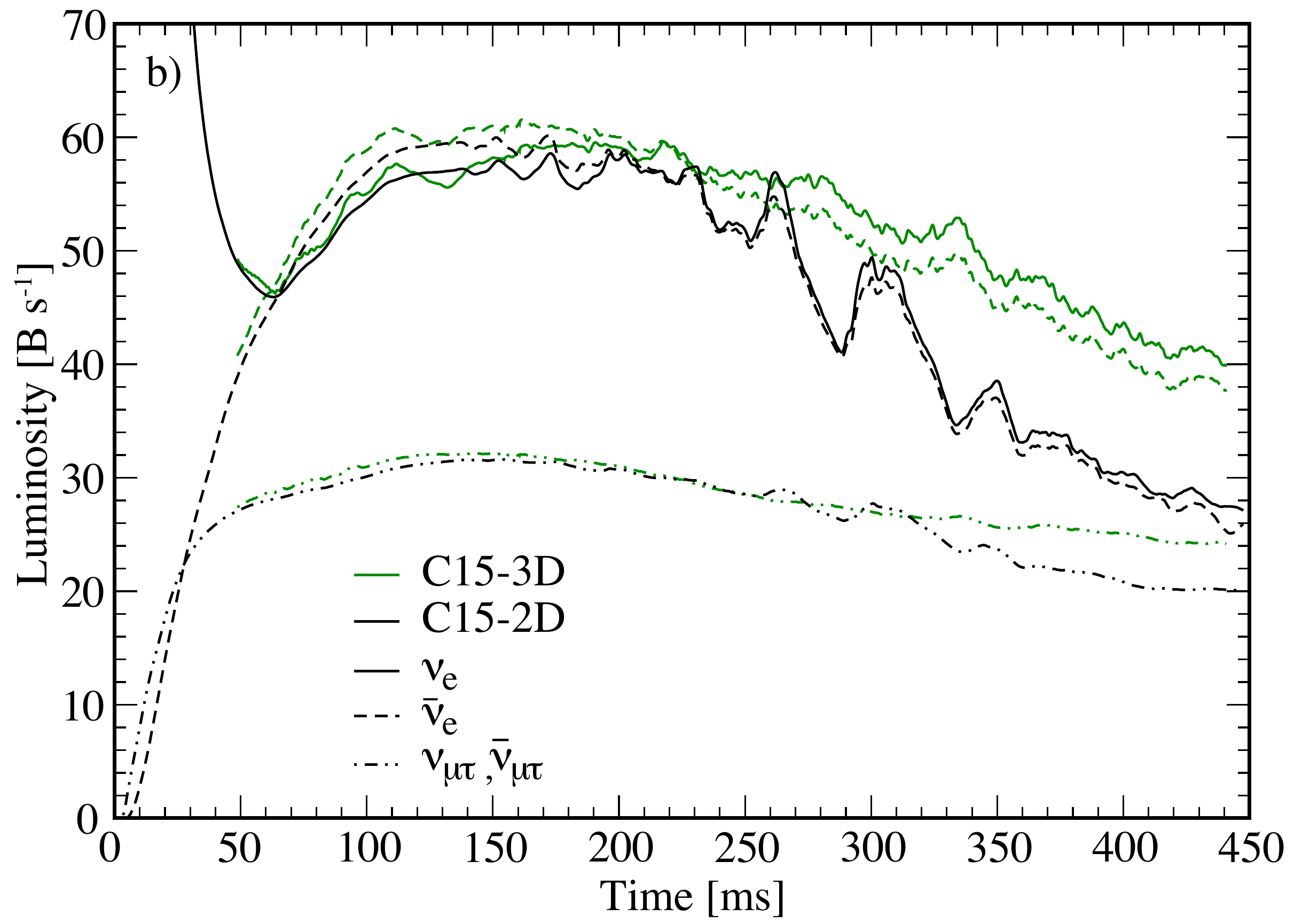}
\end{center}
\begin{center}
\includegraphics[width=.95\columnwidth]{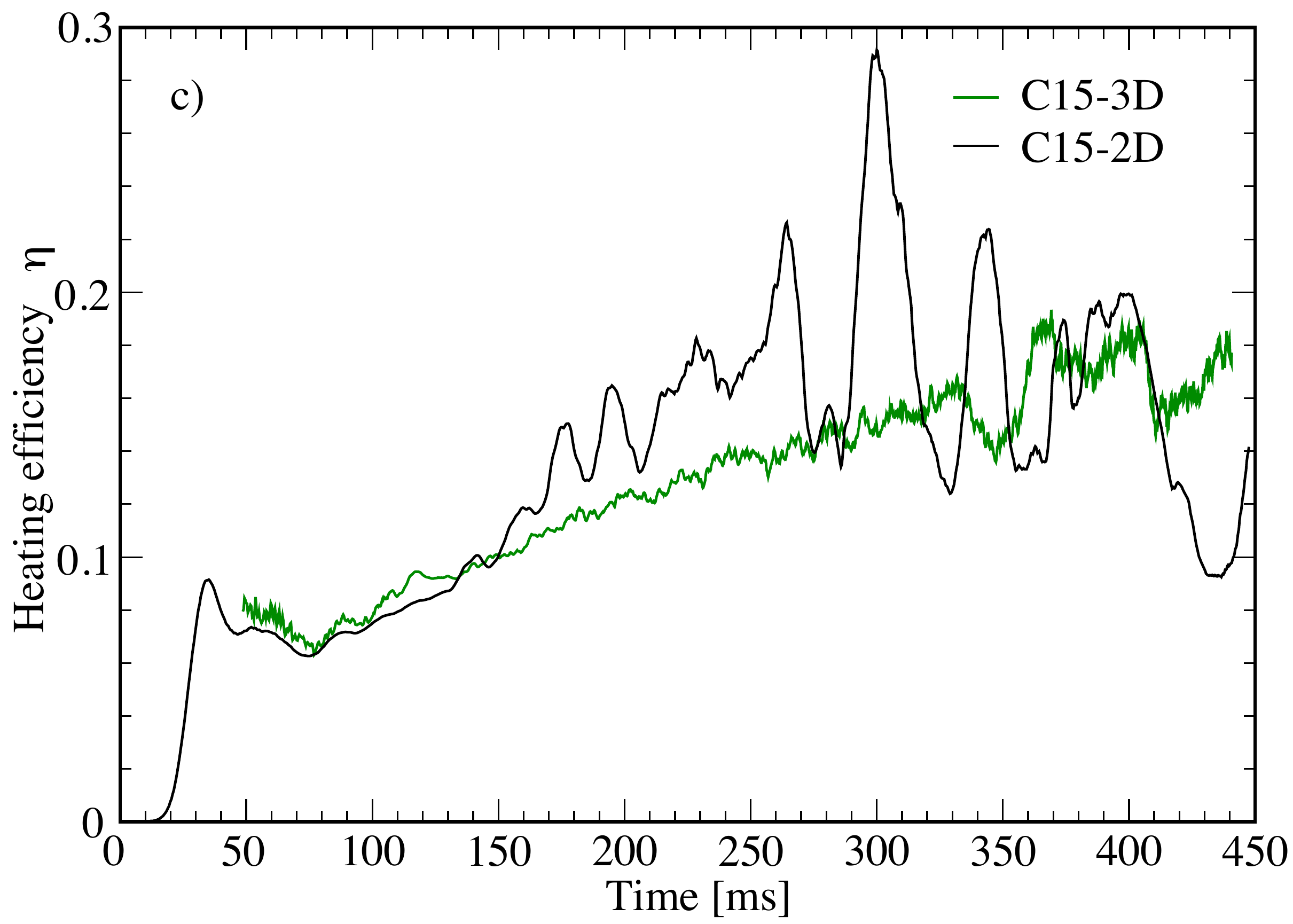}
\includegraphics[width=.95\columnwidth]{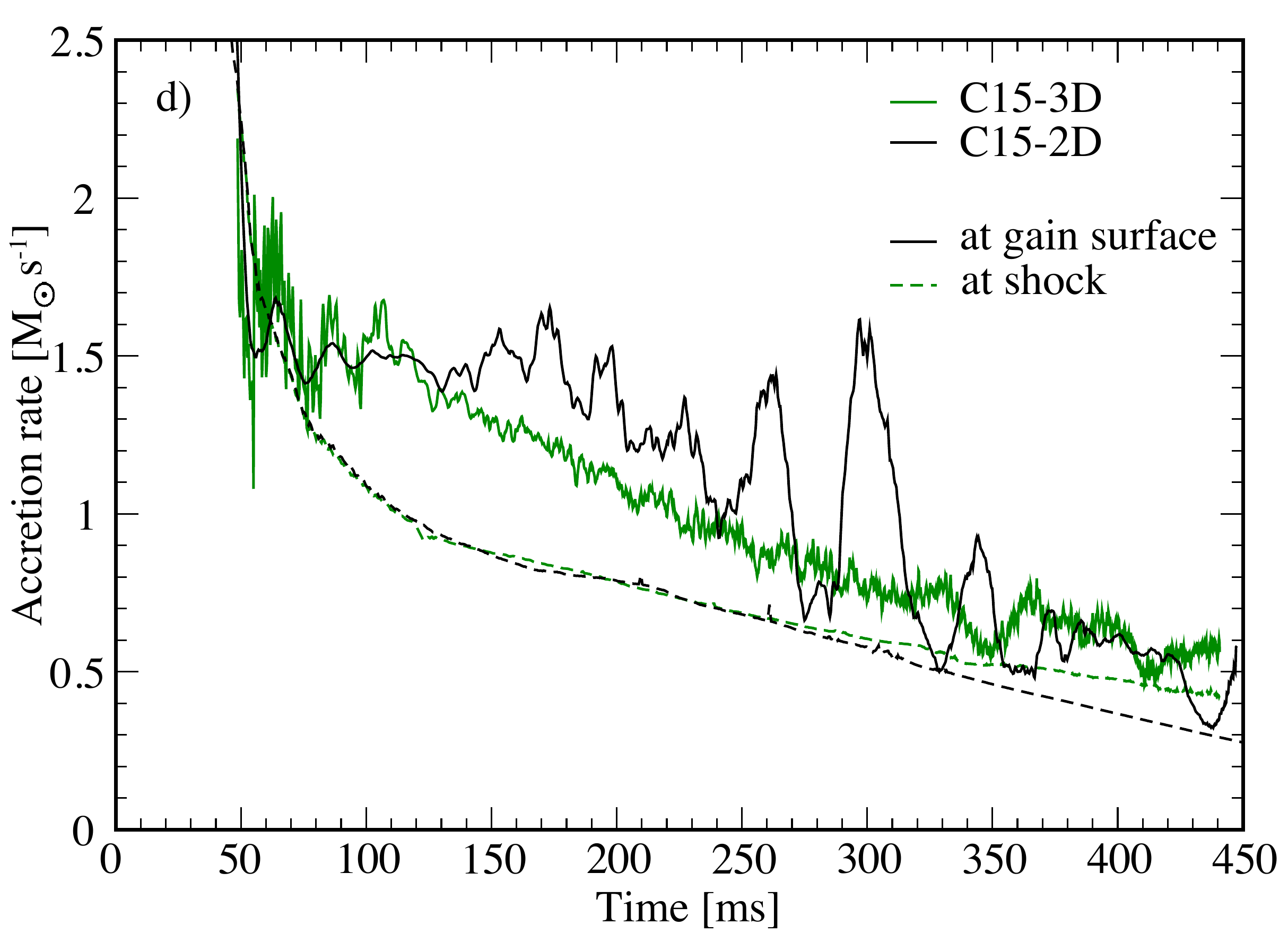}
\end{center}
\begin{center}
\includegraphics[width=.95\columnwidth]{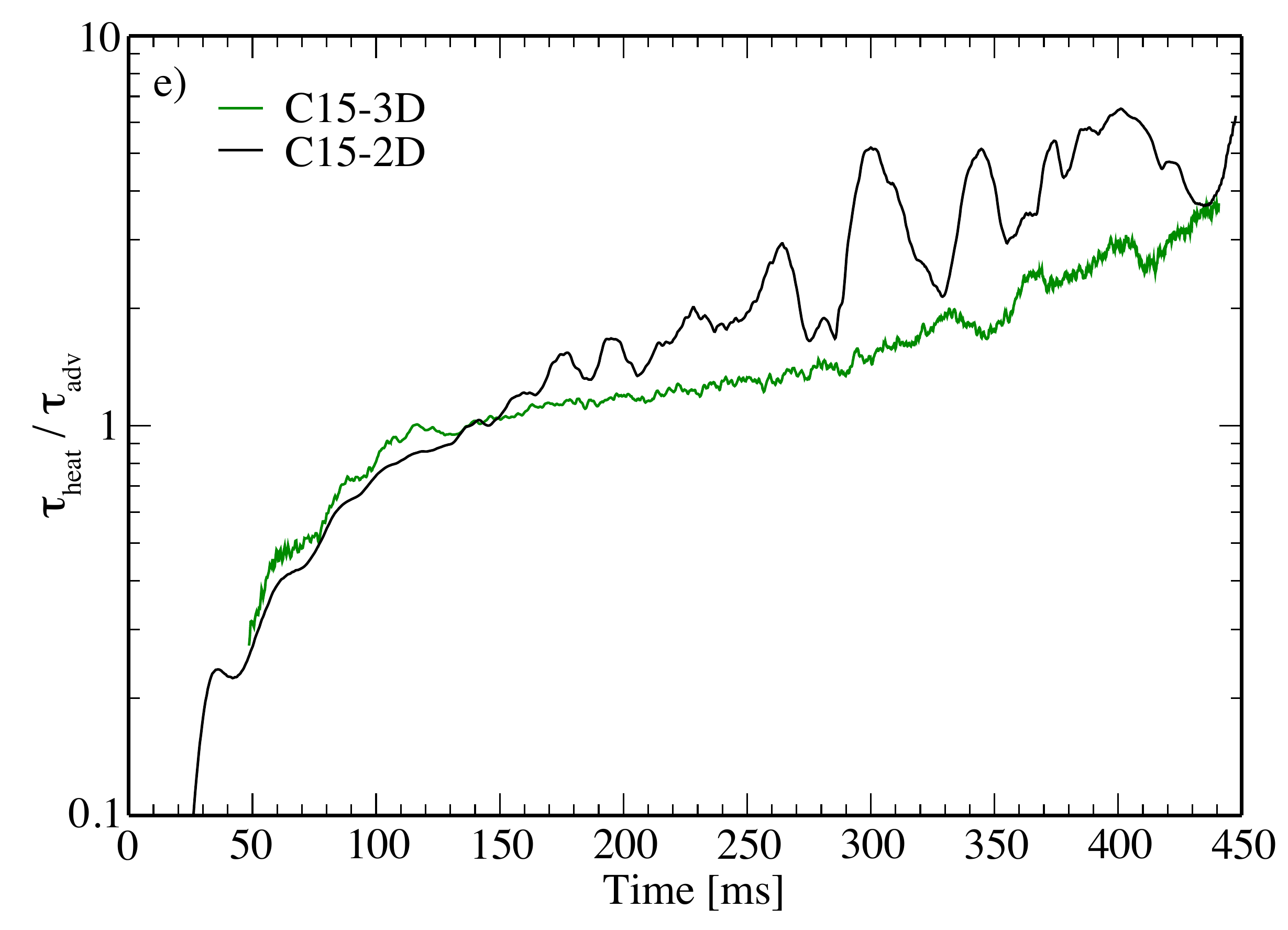}
\includegraphics[width=.95\columnwidth]{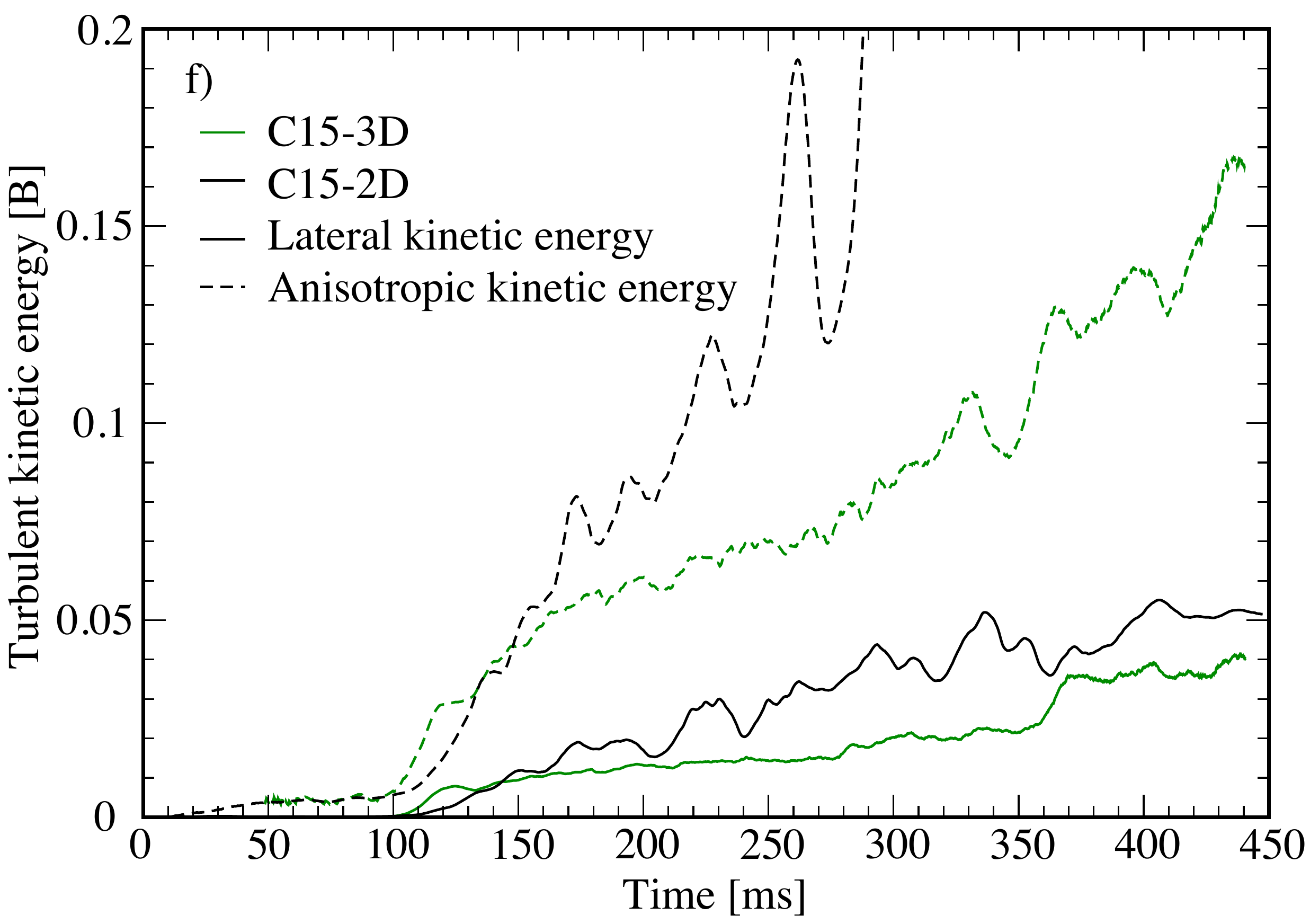}
\end{center}
\caption{\label{fig:traces}
 a) Net neutrino heating in the gain region.
 b) \nue\ (solid), \nuebar\ (dashed), and \numt\ (dash-dotted) total luminosities at 1000~km.
 c) Neutrino heating efficiencies.
 d) (inward) Accretion rates at gain radius (solid) and shock (dash-dotted).
 e) Advection--heating time scale ratio, $\tadv/\theat$.
 f) Turbulent kinetic energy.
Data for C15-2D is averaged with a 25-point boxcar ($\sim$8~ms). Plotted using colors of Figure~\ref{fig:shockradii}.
}
\end{figure*}

Neutrino heating establishes a heating region extending inward from the shock to the gain surface, where net neutrino heating transitions to net cooling.
Starting at $\approx$80~ms for both multi-D simulations, heating at the base of the gain region creates buoyantly unstable conditions, resulting in convective plumes rising against the continuing inflow.
Rising plumes begin to affect the shock surface at $\approx$95~ms for C15-3D and $\approx$105~ms for C15-2D, as seen by the separation of the minimum and maximum shock radii (Figure~\ref{fig:shockradii}; dashed lines).
Over the next $\sim$50~ms, both models become completely convective within the shocked cavity.
For C15-3D this results in a flat mean shock radius, \Rshock, that rises gradually until $\approx$280~ms.
For C15-2D, \Rshock\ oscillates and grows faster, indicating earlier shock revival and explosion.
The shock for C15-1D, which lacks multi-dimensional flows, reaches a maximum radius of $\approx$180~km at $\approx$80~ms and recedes thereafter, typical of 1D CCSN simulations.

The shock in C15-2D expands rapidly from $\approx$230~ms onward (Figure~\ref{fig:shockradii}), with the diagnostic energy\footnote{following \citetalias{BrLeHi14}, \Ediag\ is defined as the integral of the total energy (thermal, kinetic, and gravitational) in all zones of the cavity where locally positive.} \Ediag\ (Figure~\ref{fig:overview}a) simultaneously becoming positive.
\Ediag\ surpasses 0.01~B by 250~ms and grows rapidly thereafter.
For C15-3D, the first evidence of potential explosion begins with an increased growth of \Rshock\ at $\approx$280~ms,  accelerating after $\approx$350~ms, as the largest buoyant plume expands, leading to a small, but growing \Ediag.

The explosion is clearly more energetic in C15-2D at all times (Figure~\ref{fig:overview}a).
We evaluate the growth of \Ediag\ over a common period beginning when \Rshock\ exceeds 500~km and ending 45~ms later.
For C15-3D, \Rshock\ passes 500~km at 393~ms when \Ediag\ is 0.034~B, which grows to 0.067~B at 438~ms when \Rshock\ is 735~km.
For C15-2D, \Rshock\ exceeds 500~km at 278~ms when \Ediag\ is 0.041~B, which grows to 0.147~B at 323~ms when \Rshock\ reaches  900~km.
Over this 45~ms comparison period, the \Ediag\ growth rate is 0.73~\Bethes\ for C15-3D and 2.35~\Bethes\ for C15-2D, and the growth rate of \Rshock\ in C15-2D is nearly double that of C15-3D.
These indicate stronger growth of explosion in 2D.
Further assessment of \Ediag\ growth requires significantly longer simulations: full saturation of \Ediag\ took $\sim$1 second to reach $\sim$1~B for the equivalent 2D model (B15-WH07) of \citetalias{BrLeHi14}.

The mass in the shocked cavity for C15-2D diverges from that of C15-3D at $\approx$150~ms (Figure~\ref{fig:overview}b; solid) and trends strongly upward by $\approx$250~ms corresponding to rapid growth in \Rshock\ and \Ediag.
Similarly, the mass in the gain region (dashed) grows from $\approx$220~ms onward.
For C15-3D, the turn-around in the mass curves is shallower, later, and occurs with less mass in the shocked cavity and gain region --- all factors that correlate with weaker (or delayed) explosions.

\section{Differences between 2D and 3D}
\label{sec:2d3d}

With successful explosions manifest in both 2D and 3D, the leading question is: ``Why do axisymmetric models proceed more rapidly (and more forcefully) to explosion than 3D counterparts?''

From 50~ms to 150~ms, C15-3D shows larger total heating in the gain region (Figure~\ref{fig:traces}a), arising from greater heating efficiency (Figure~\ref{fig:traces}c; \heateff; luminosity divided by net heating rate) and neutrino luminosities (Figure~\ref{fig:traces}b), though mass accretion at the gain radius (Figure~\ref{fig:traces}d; \accgain) is similar, resulting in larger \Rshock\ than in C15-2D.
At $\approx$150~ms, the ratio of advection and heating time scales ($\tadv/\theat$; Figure~\ref{fig:traces}e), defined in \citetalias{BrLeHi14},  grows past unity in both simulations, signaling the potential for thermal runaway \citep{BuJaRa06}.
In C15-2D, $\tadv/\theat$ grows more rapidly, with large excursions driven by the oscillation of the shock along the pole.
During this epoch, \accgain\ is larger in C15-2D, with large, positive excursions, correlated with favorable increases in the luminosities, \heateff, and $\tadv/\theat$, continuing through C15-2D shock revival ($\approx$250~ms).
This favors earlier development of explosion in C15-2D, even though the luminosities remain higher in C15-3D.
For C15-2D, luminosities and heating drop with accretion rate after shock revival at $\approx$250~ms, while both remain noticeably higher in C15-3D.
These measures of accretion, luminosity, and heating are generally consistent with the early development of explosion in C15-2D.

\begin{figure*}
\begin{center}
  \includegraphics[height=2.45in,clip]{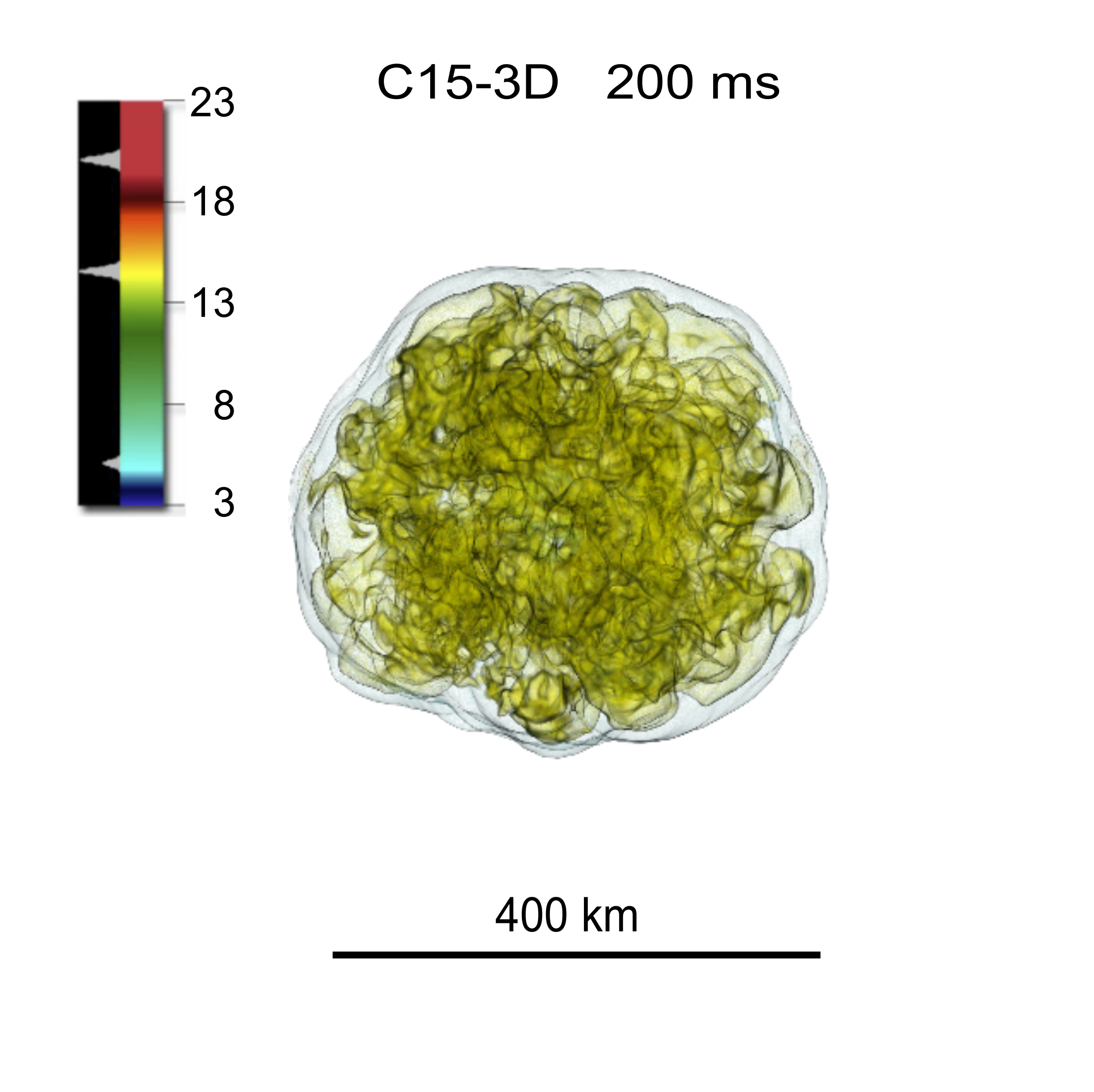}
  \includegraphics[height=2.45in,clip]{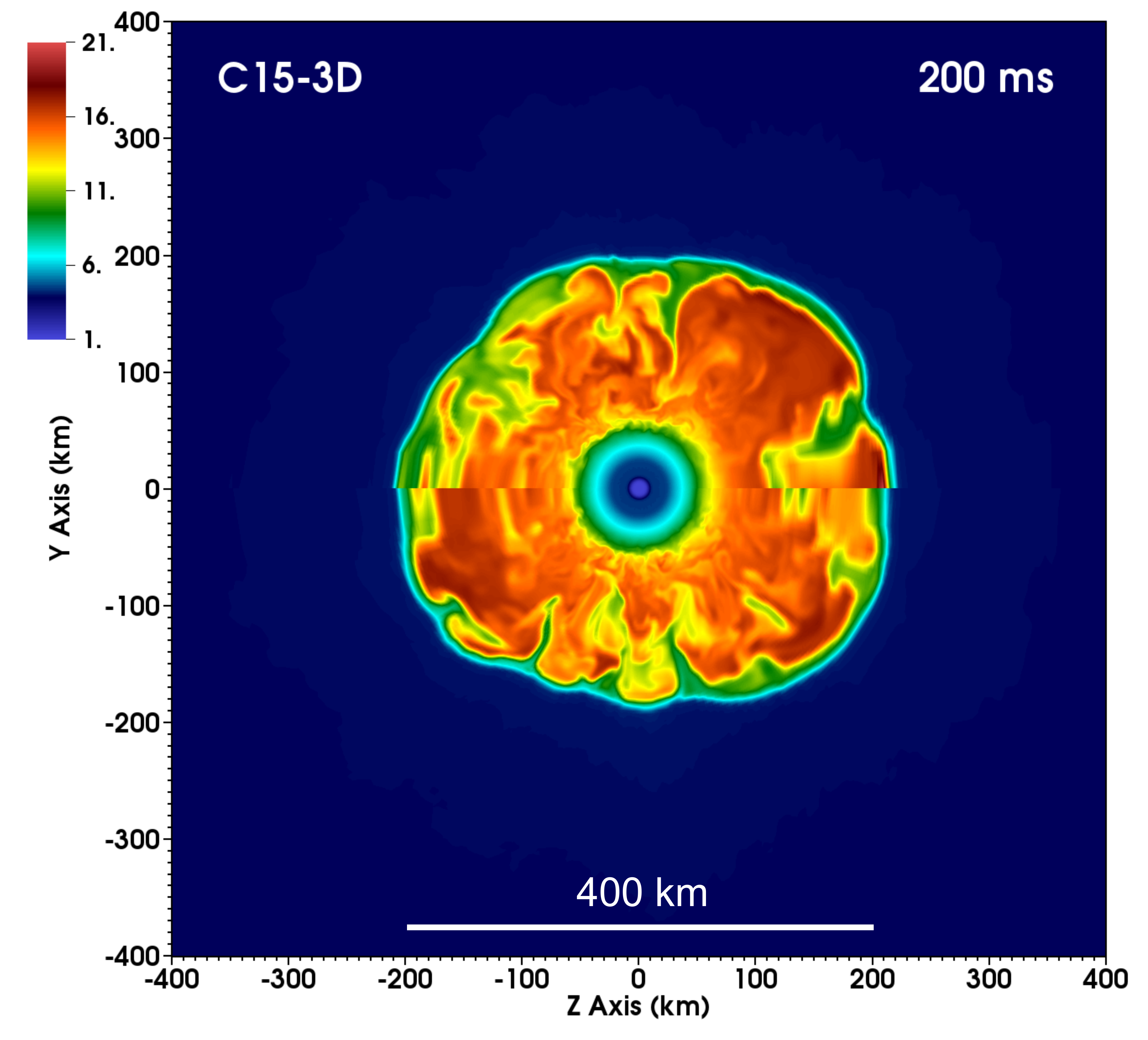}
  \includegraphics[height=2.45in,clip]{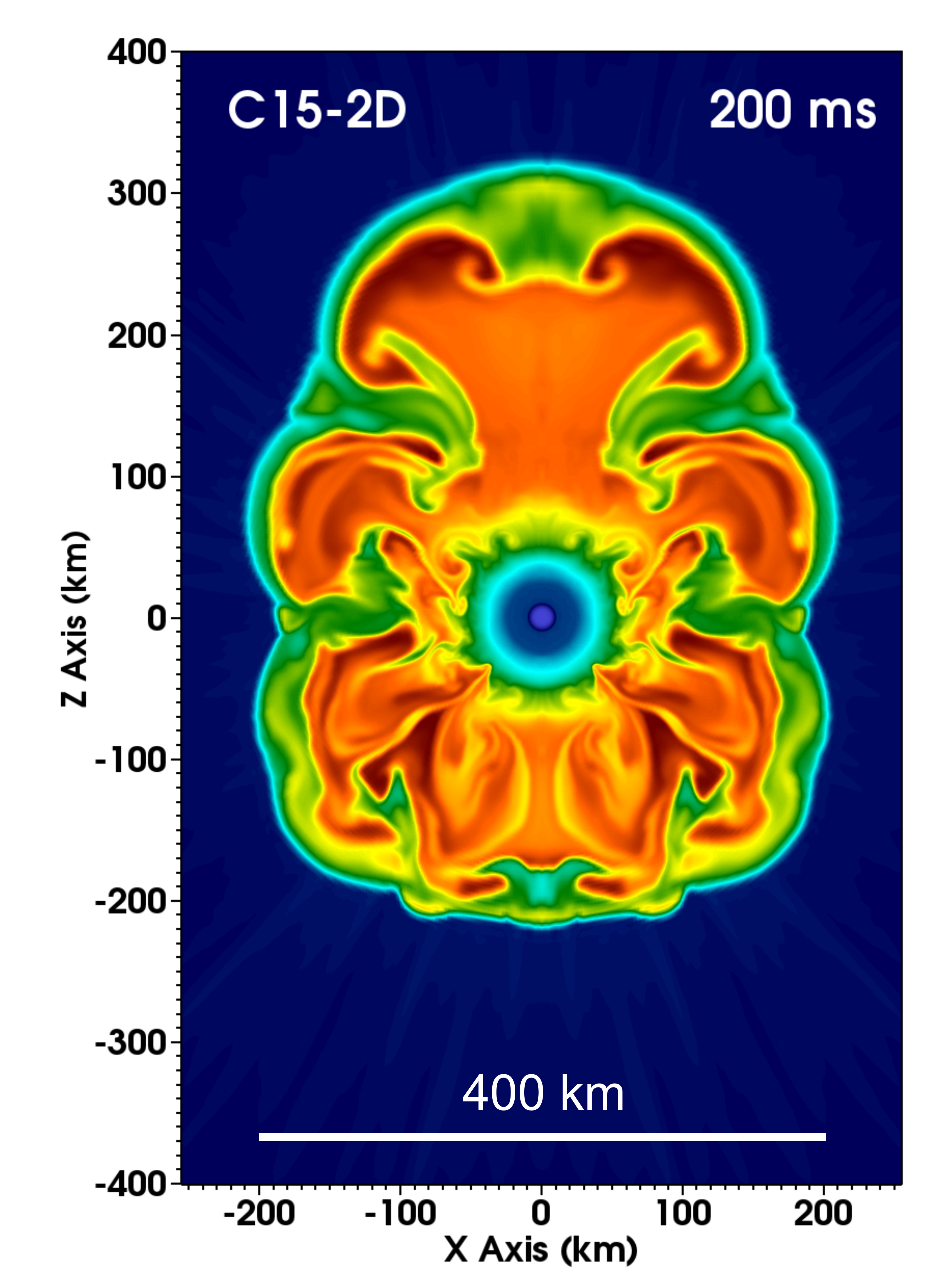}
\end{center}
\begin{center}
  \includegraphics[height=2.45in,clip]{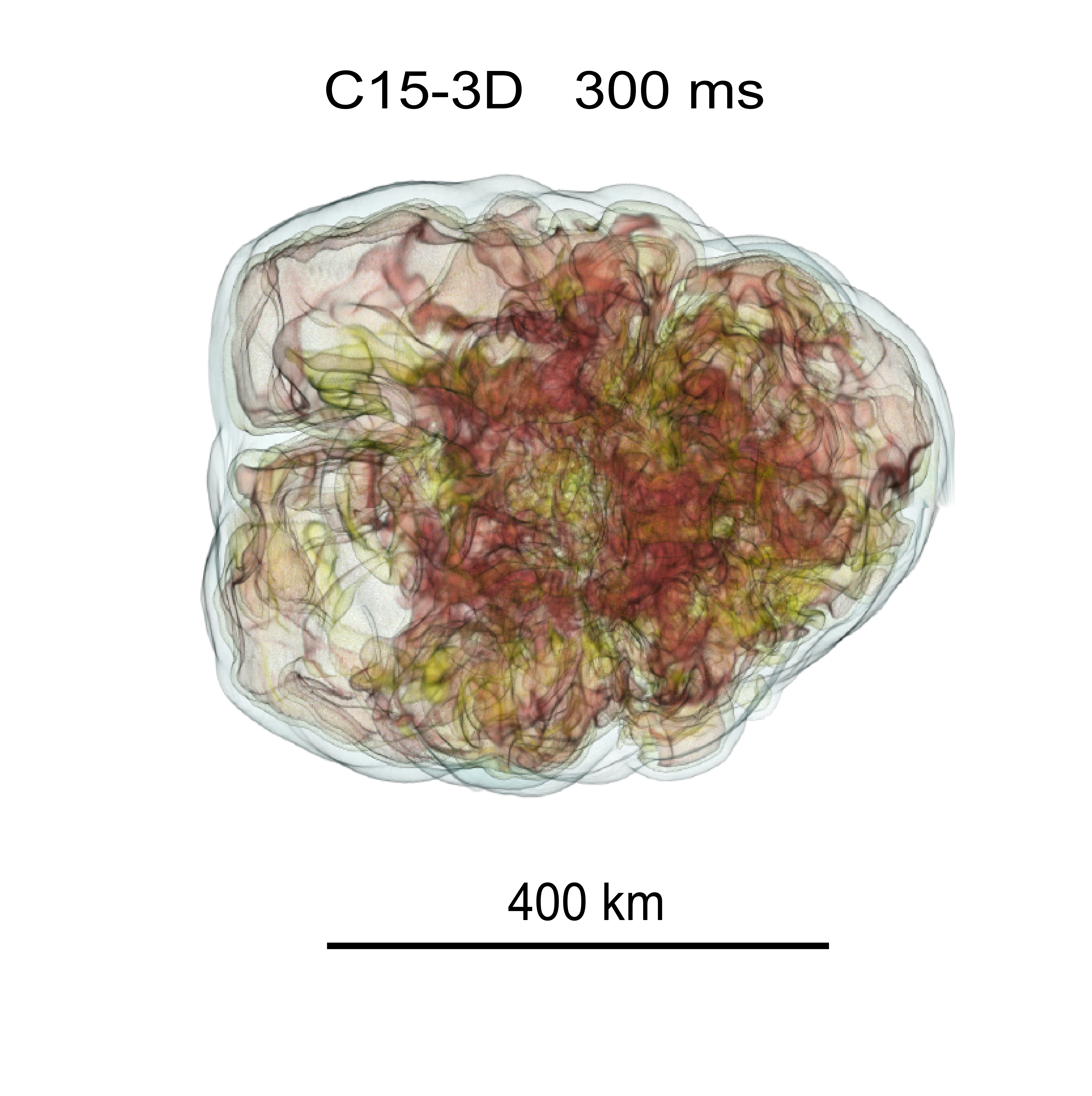}
  \includegraphics[height=2.45in,clip]{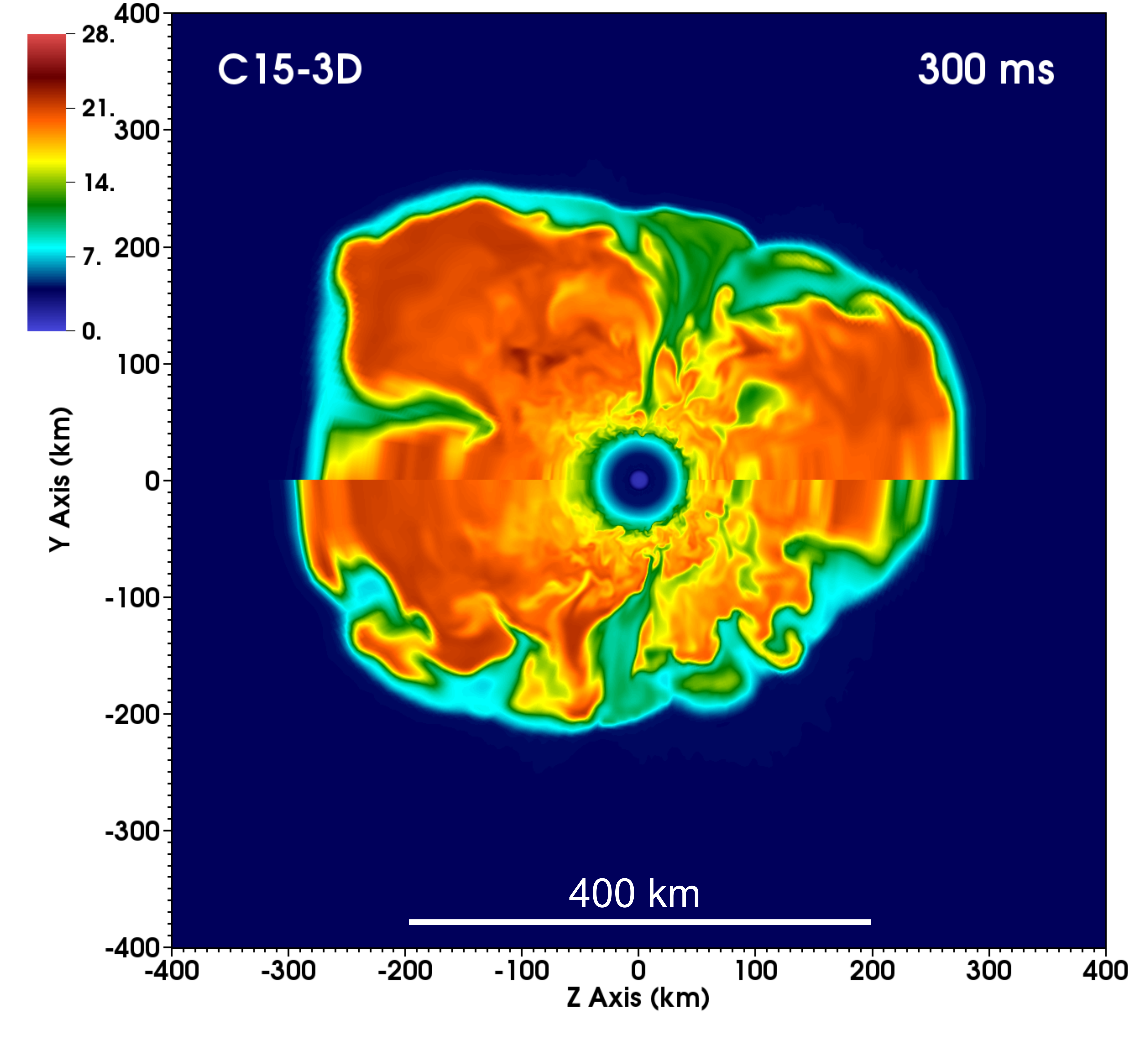}
  \includegraphics[height=2.45in,clip]{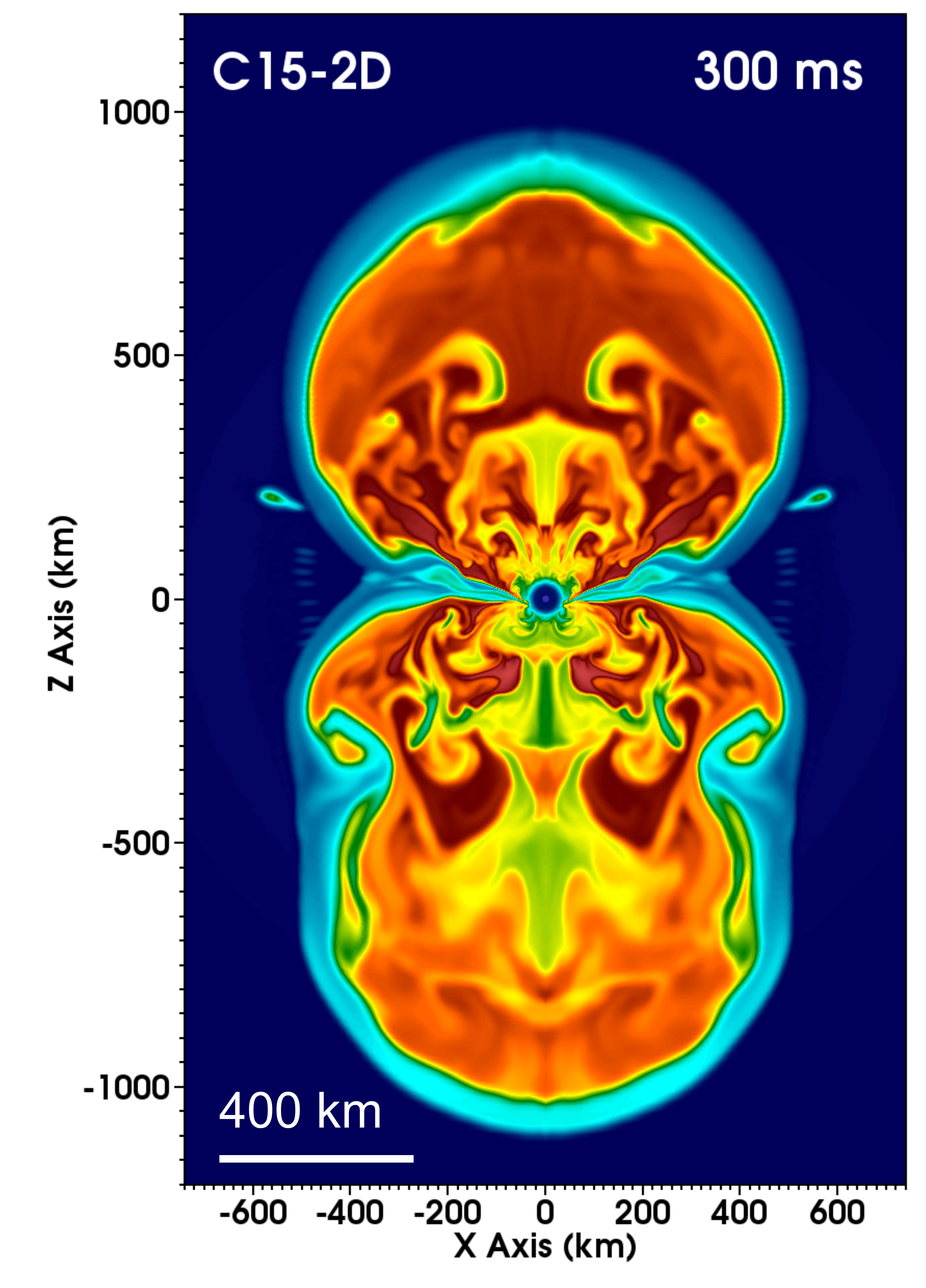}
\end{center}
\begin{center}
  \includegraphics[height=2.45in,clip]{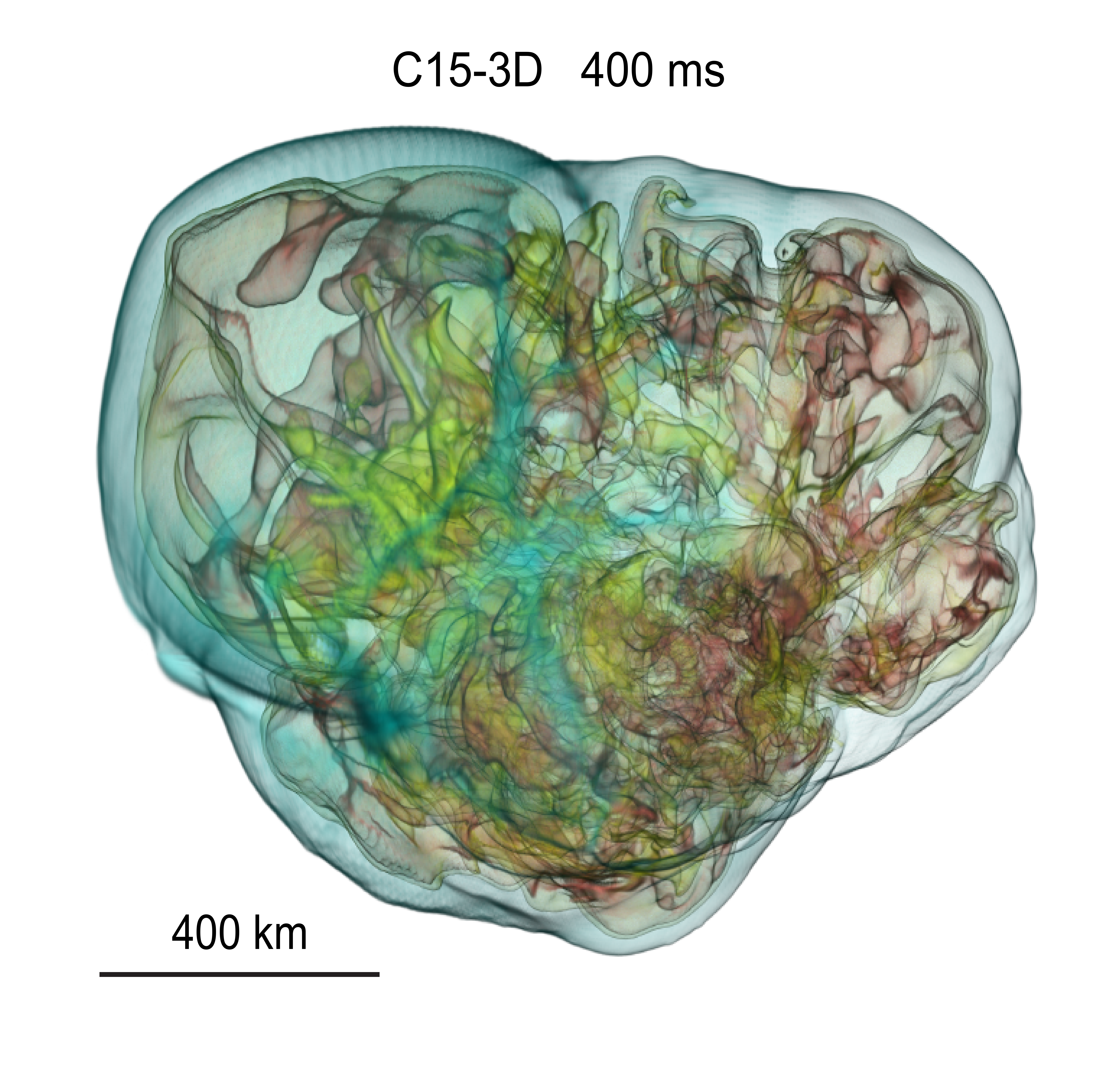}
  \includegraphics[height=2.45in,clip]{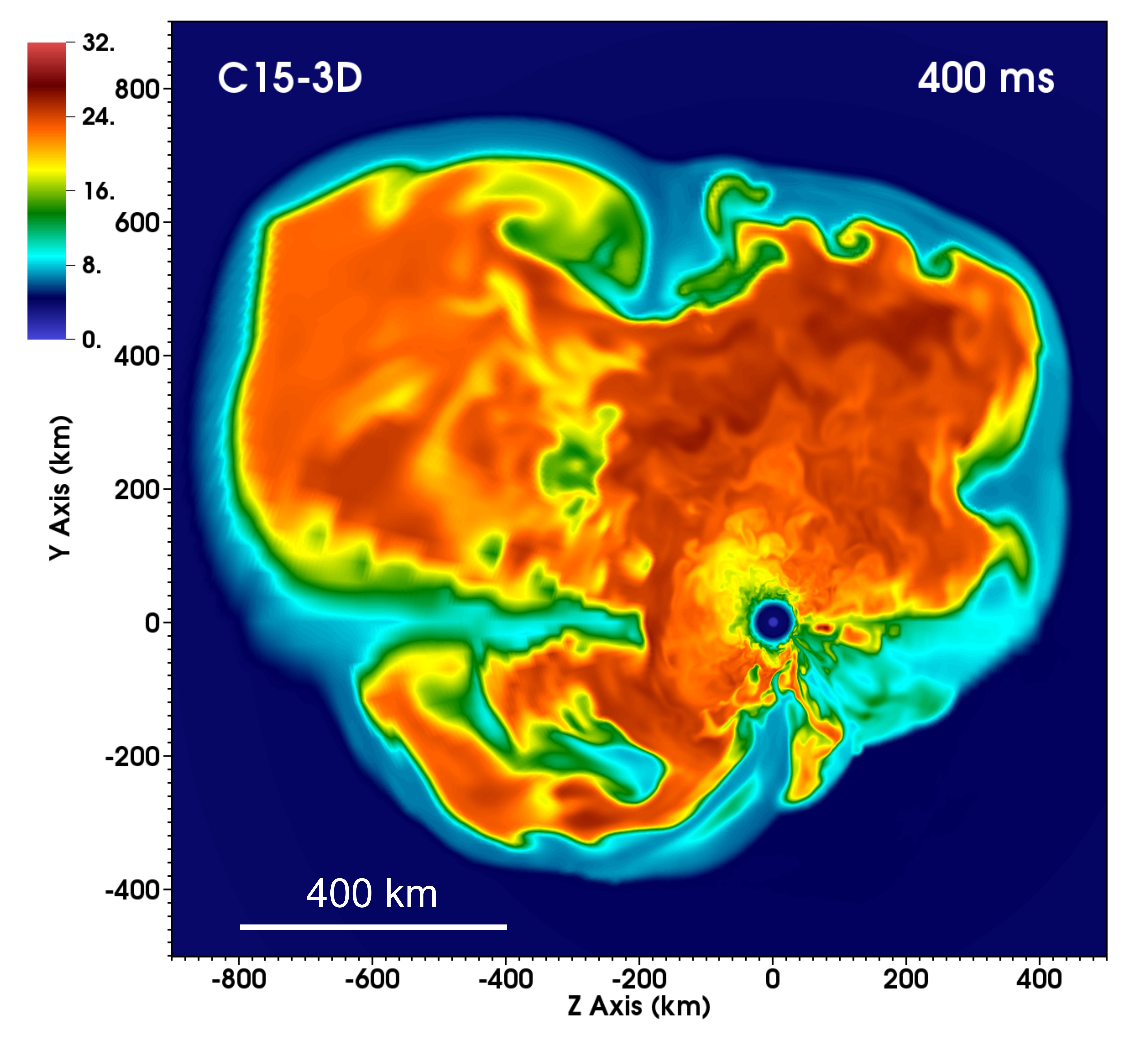}
  \includegraphics[height=2.45in,clip]{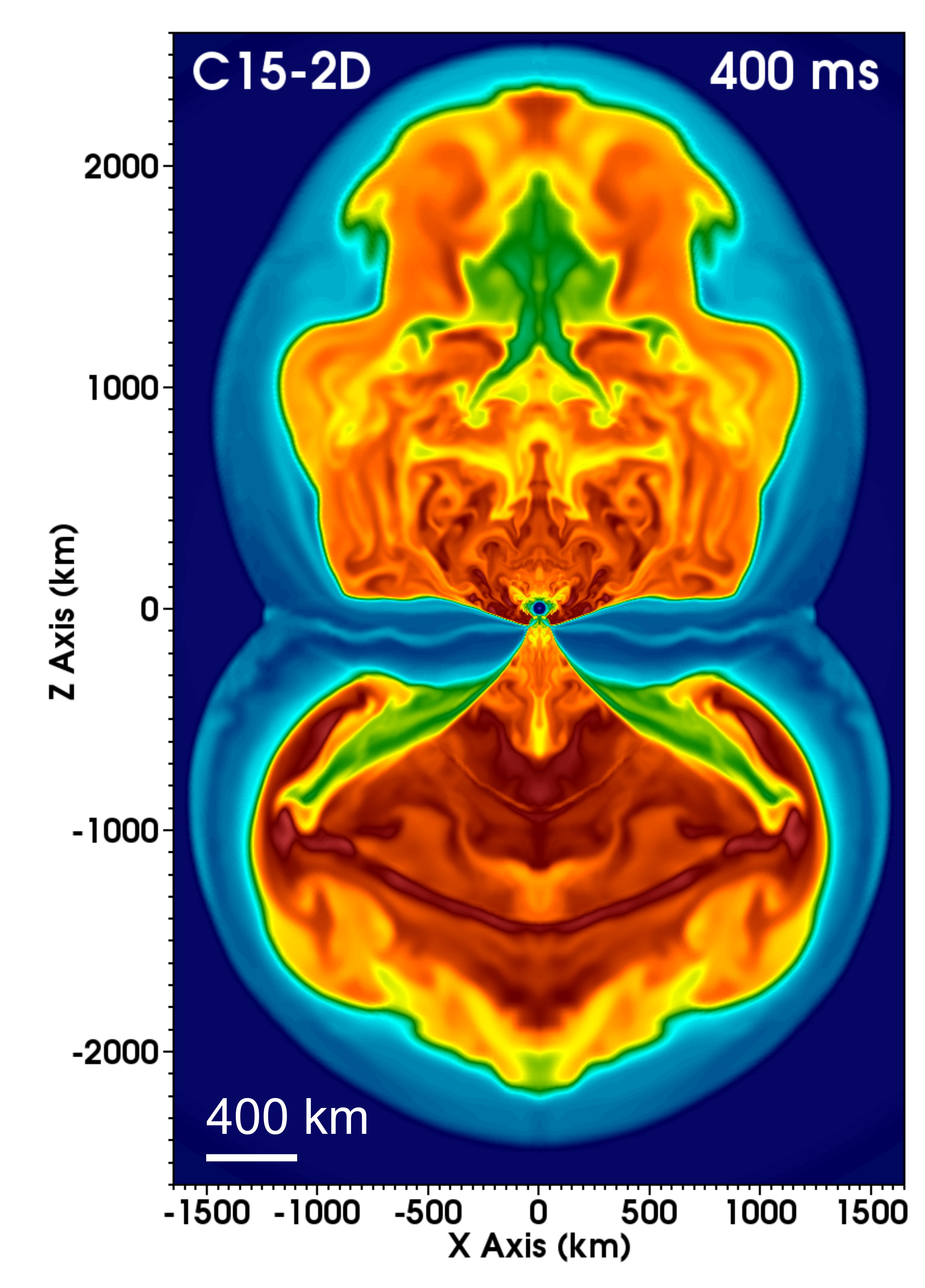}
\end{center}
\caption{\label{fig:volume}
Specific entropy (\kbbar) at 200, 300, and 400~ms with 400-km scale bars in each panel.
Column a (left): Volume rendering for C15-3D using a fixed transfer function, highlighting rising plumes.
Column b (center): Polar slice through C15-3D, aligned with Column~a. In upper two panels (200 and 300~ms), the 180\degree\ $\phi$-shift between upper and lower halves is exaggerated by the 8.5\degree\ zone at the pole. 400-ms panel shows effect of transition to $\phi$-averaging at pole. 
Column c (right): Entropy in a polar slice through C15-2D with color scale matching Column~b at each epoch.  
Animated version of Column~c available at \url{ChimeraSN.org}.
}
\end{figure*}

Recent work has examined the role turbulence can have in supporting the shock leading up to explosion \citep[e.g.,][]{MuMe11,CoOt15,AbOtRa14}.
In Figure~\ref{fig:traces}f we show the kinetic energy associated with turbulence in the gain region.
Turbulent kinetic energy
\begin{equation}
\Eturb{X} = \int_{\rm gain} \frac{1}{2}\rho |\vturb{X}|^2 \, dV,
\end{equation}
is defined for velocities, \vturb{X}, integrated over the gain region.
Lateral turbulent energy, \Eturb{lat}, is computed by setting radial velocity $v_r\equiv0$ in defining \vturb{lat}.
Anisotropic turbulent energy \Eturb{an}, is computed by removing the radial-shell mean $\left< v_r \right>$ from \vturb{an}, $v_r^{\rm an} \equiv v_r - \left< v_r \right>$.
The solid lines show the growth of \Eturb{lat}, which begins growing at $\approx$100~ms, the onset of non-radial motions from convection, and continues afterward.
\Eturb{an} (dashed lines) is approximately four-fold larger in both simulations prior to shock revival.
Both measures are larger for C15-2D than for C15-3D, consistent with \citet{CoOt15}, who posited that stronger turbulent pressure aids the development of explosions in 2D simulations.
It is important to note that while \Eturb{} measures the kinetic energy of disordered flow, the relevant driver is convection driven by neutrino heating.
This is especially important for \Eturb{an} where accretion streams and rising plumes are both deviations from the mean radial flow at large scales.

For multi-dimensional models (including those discussed above), there is a pre-explosion state with convective plumes rising to and distorting the shock.
The flow across the accretion shock is diverted by shock geometry toward the local shock minima and then into accretion streams between the plumes.
C15-3D, like previously reported 3D simulations, initially shows a large number of small plumes (Figure~\ref{fig:volume}a,b), whereas C15-2D (and most other 2D simulations) shows rapid development toward only a few large plumes (Figure~\ref{fig:volume}c).
As in \citetalias{BrLeHi14}, the primary polar plumes in C15-2D oscillate along the symmetry axis, in a manner consistent with the SASI, while neutrino-heated material continues to flow into plumes from the bottom of the gain region, quickly triggering shock revival.
In C15-3D, initially small rising plumes are pushed back by ram pressure at the shock and lack the persistence of the larger polar plumes in 2D models (see animated version of Figure~\ref{fig:volume}a).

\begin{figure*}
  \begin{center}
\includegraphics[width=.6\columnwidth,clip]{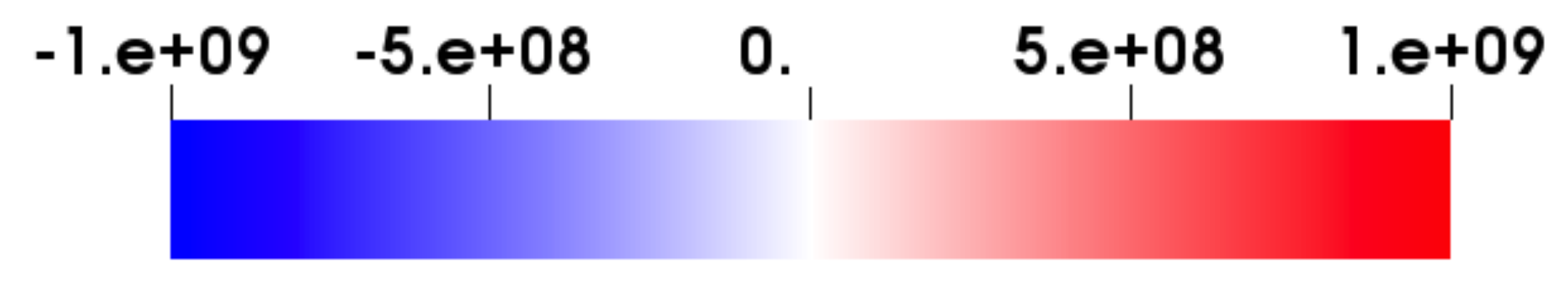}
  \end{center}
  \begin{center}
\includegraphics[width=\columnwidth,clip]{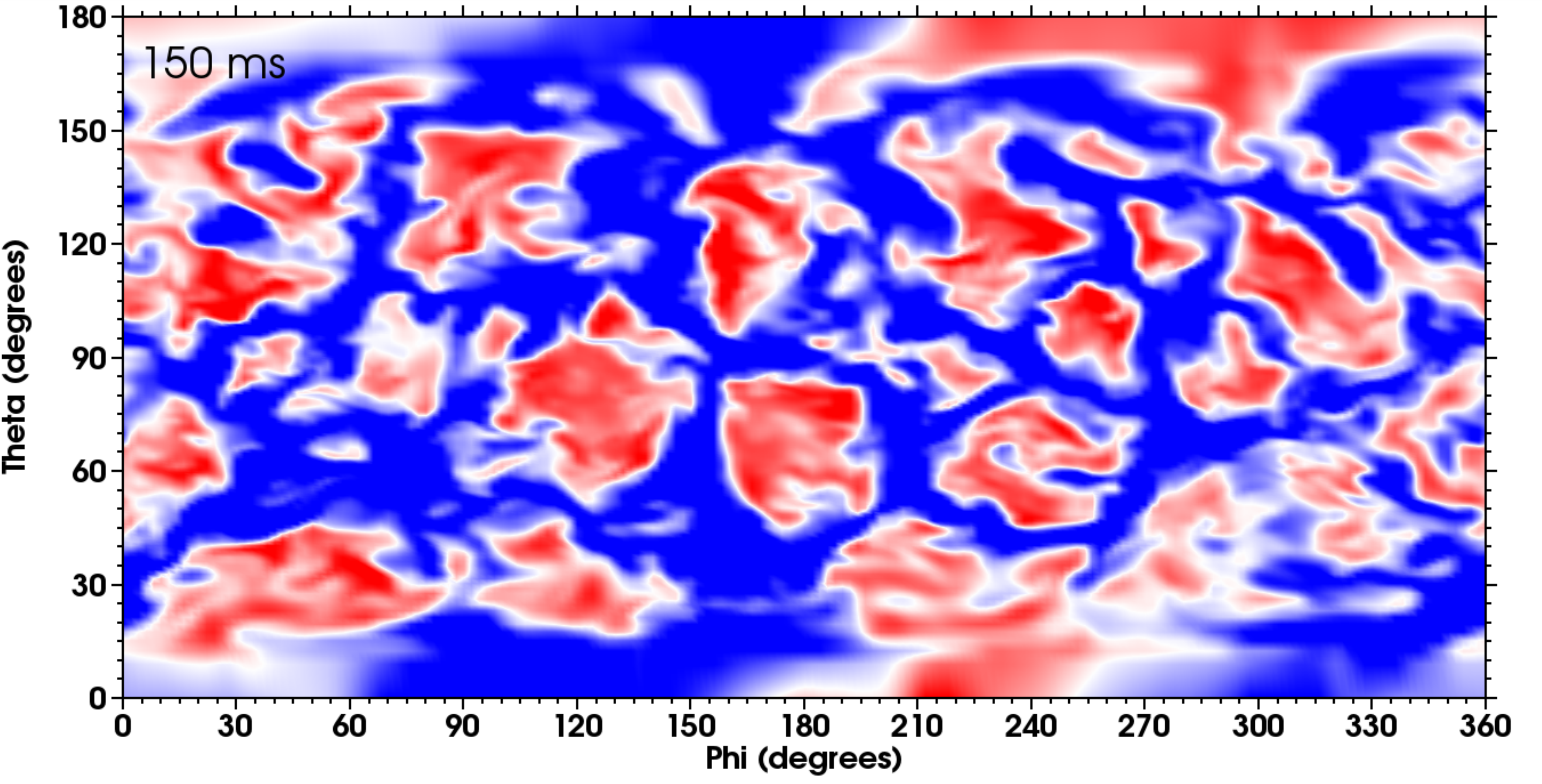}
\includegraphics[width=\columnwidth,clip]{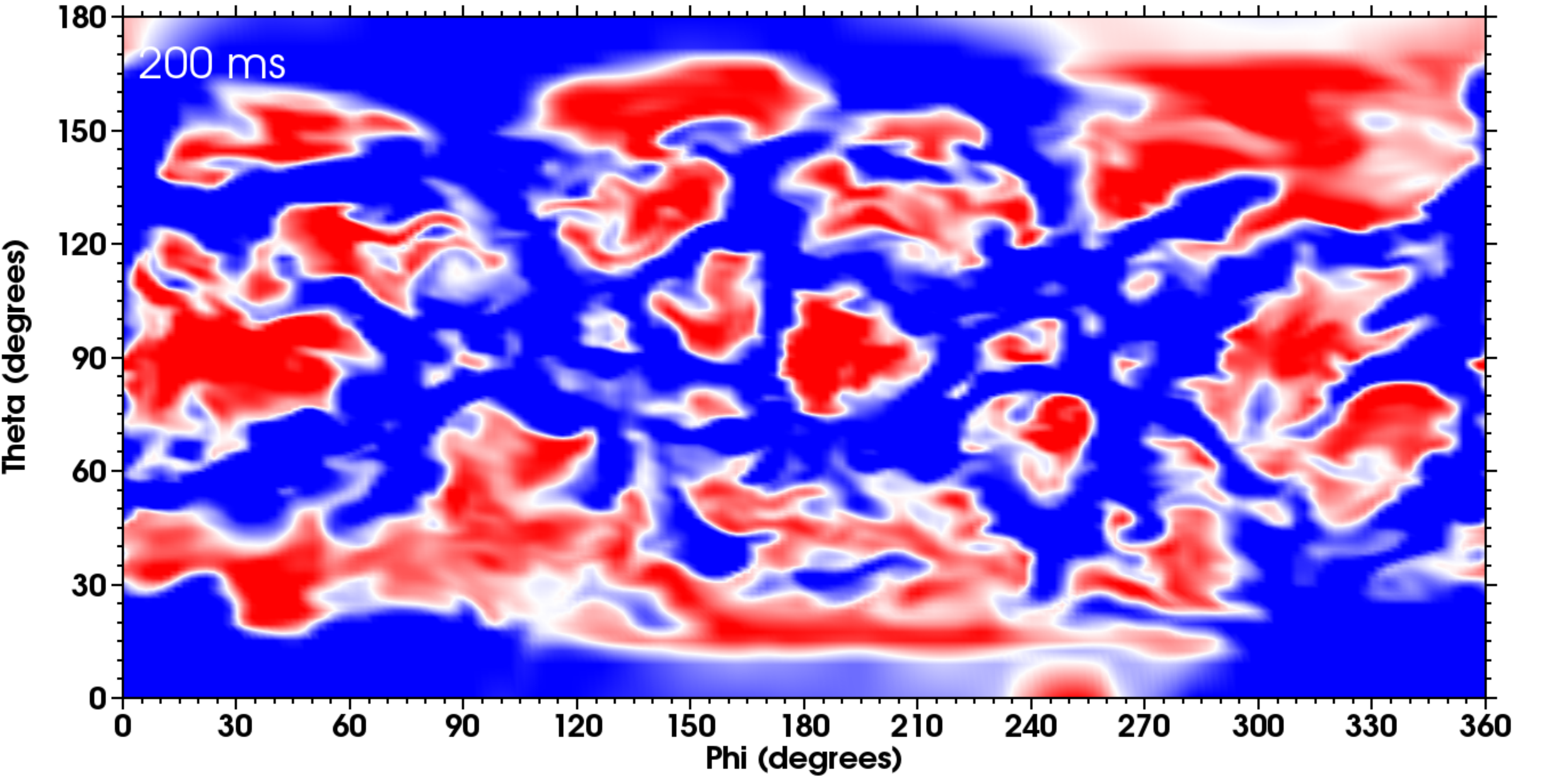}
  \end{center}
  \begin{center}
\includegraphics[width=\columnwidth,clip]{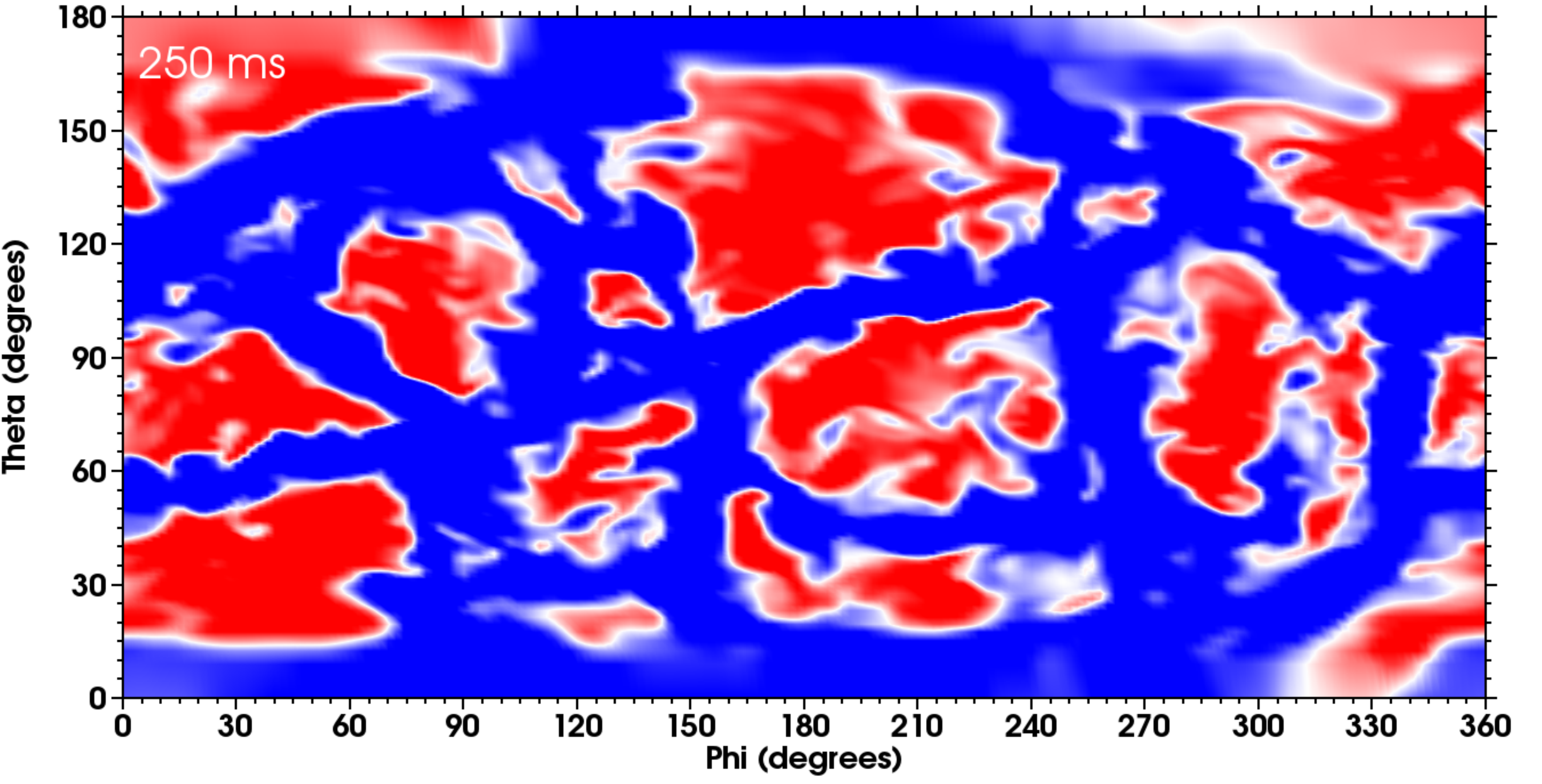}
\includegraphics[width=\columnwidth,clip]{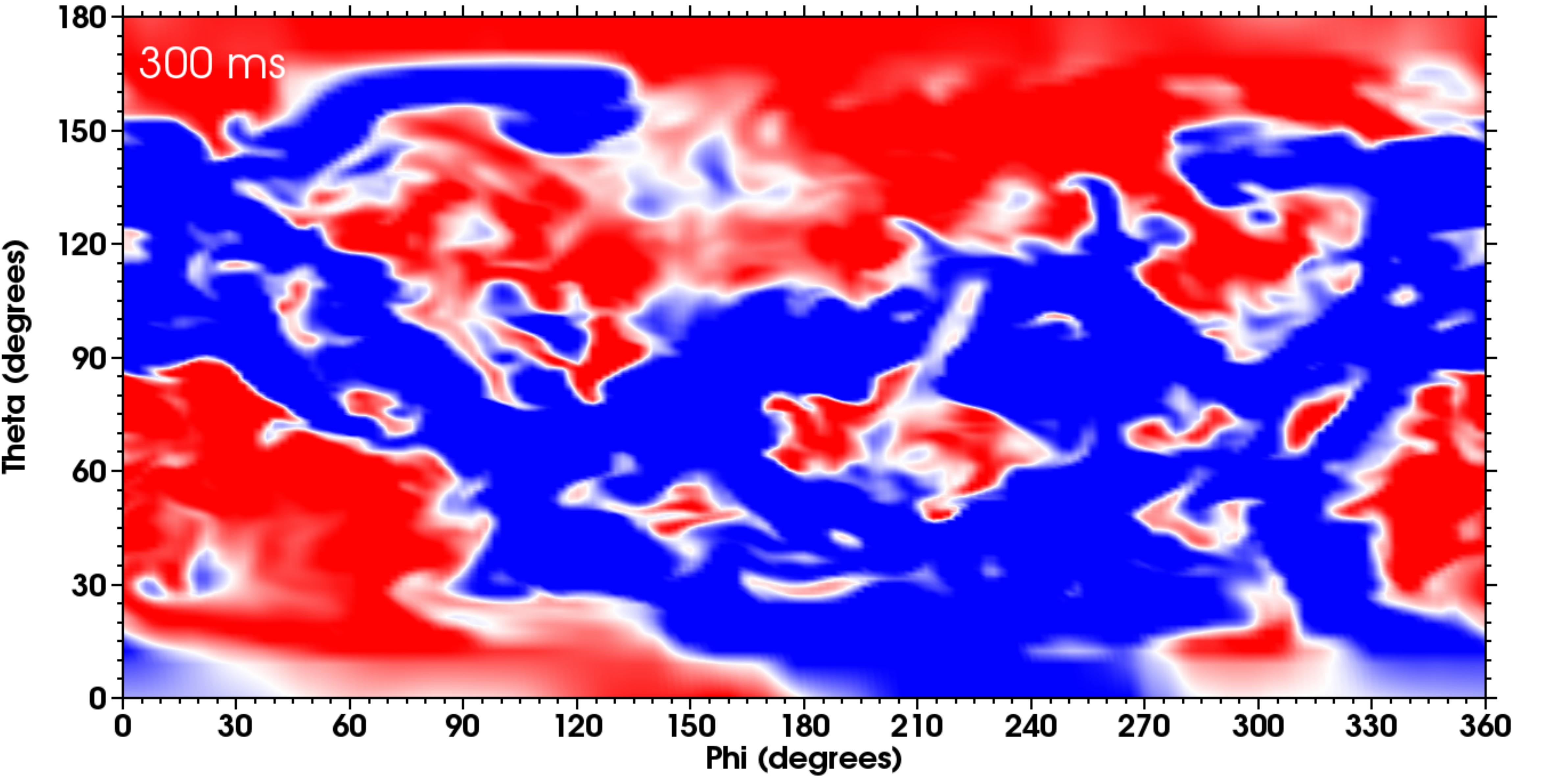}
  \end{center}
  \begin{center}
\includegraphics[width=\columnwidth,clip]{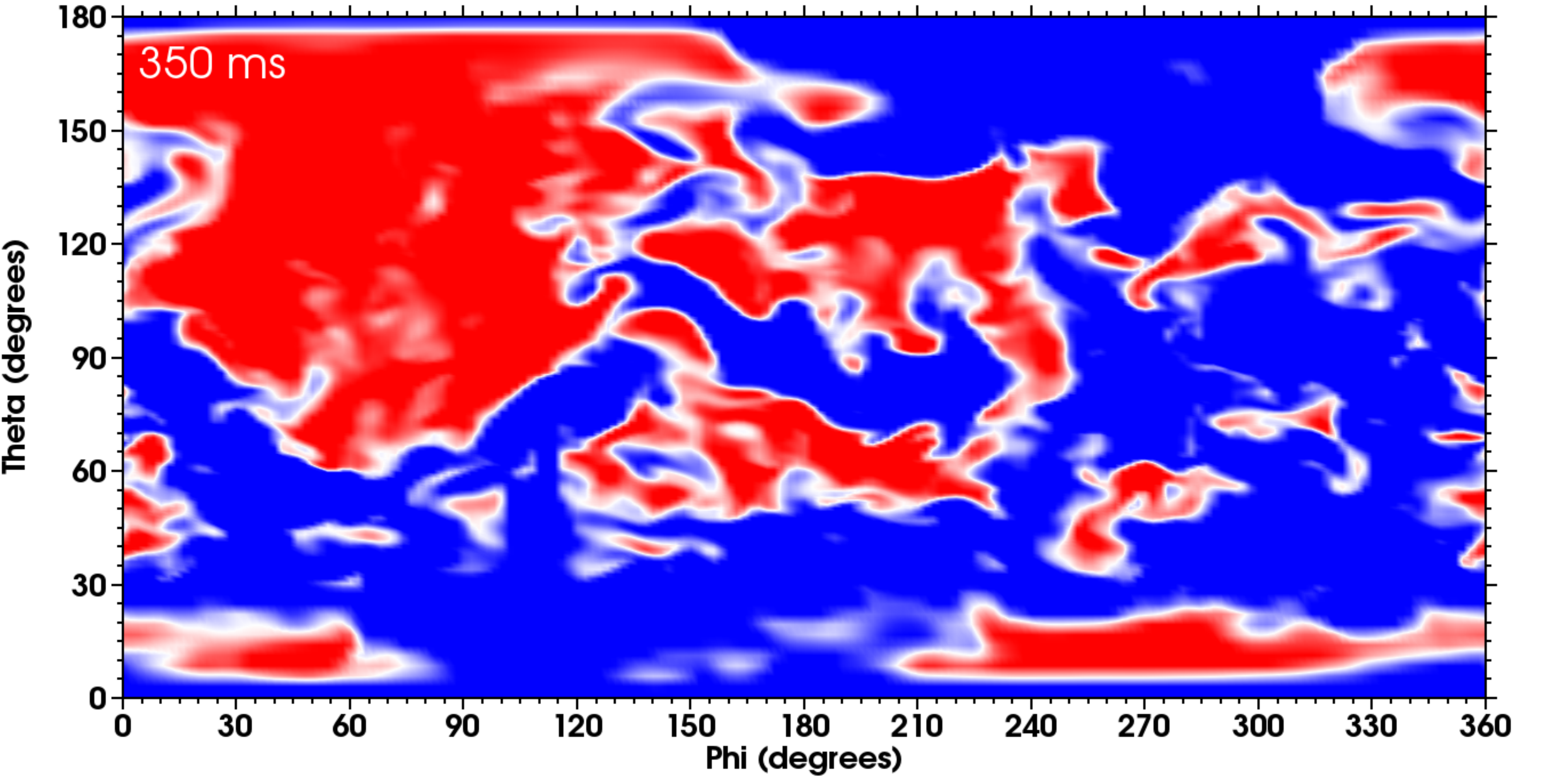}
\includegraphics[width=\columnwidth,clip]{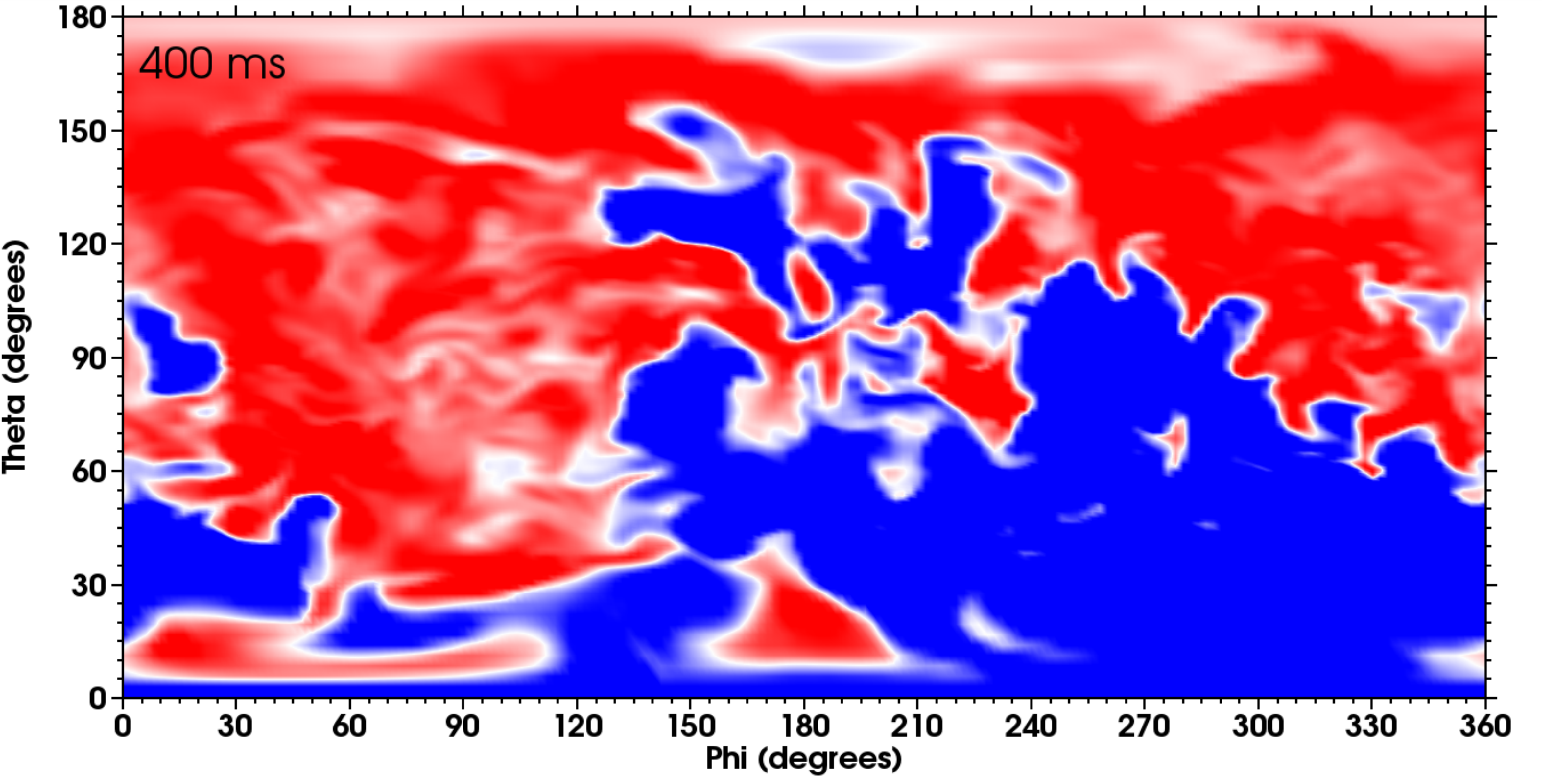}
  \end{center}
\caption{\label{fig:plumeslice}
150-km slice of radial velocity in cm~s$^{-1}$, showing development of plumes. Upflows (downflows) are red (blue) saturated at +(-)10000~\kmps.
}
\end{figure*}

As C15-3D progresses toward explosion, the angular scale of the plumes slowly grows and, eventually, the largest of these plumes expands continuously and drives shock revival.
The growth of plume scale can be seen at the surface (animation of Figure~\ref{fig:volume}a) and deeper (150~km; Figure~\ref{fig:plumeslice}). 
The growth of the plume scales clearly precedes shock revival at $\approx$350~ms, when the shock is clearly expanding and upflow at 150~km is dominated by a single large plume with a few smaller plumes.
Parameterized simulations of \citet{Fern15} show mergers of plumes and those of \citet{HaPlOd14} also found decreasing plume numbers preceding, and after, shock revival.
In 2D (including C15-2D), the axis-focused SASI and the tendency for convective vortices to merge ensures that the `large-plume-state' is reached earlier (by 150~ms in C15-2D).
In 3D this state takes time to develop and appears to delay the explosion.

We have not yet completed a comprehensive analysis of the driver of plume-scale growth, but some observations from our simulation are revealing.
As plumes grow in 3D simulations, the associated accretion streams are displaced farther from the central axis of the plume and are therefore less likely to interrupt the inflow of additional buoyant material, providing a `self-shielding' effect that permits further growth.
Additionally, the curvature of the shock surface relative to the radial inflow reduces the effective ram pressure normal to the shock surface at locations farther from the plume's axis and thus favors lateral expansion of the plume.
Together, these effects could drive the growth of larger plumes, which eventually are able to expand continuously, with material injected from the bottom of the gain region as in 2D simulations.

The effectiveness of this mechanism may be seen in the growth of the plume in the upper-left corner of the animated Figure~\ref{fig:volume}a, which precedes the onset of explosion in 3D.
The growth of this plume (also visible in the 200 and 300~ms panels of Figure~\ref{fig:volume}a,b) eventually covers a significant portion of the shocked volume and shock surface solid angle.
The lateral expansion of this plume diverts the accretion flow of material striking it farther and farther from its axis, which contributes to the formation of a strong accretion region on the opposite side of the proto-NS at $\approx$350~ms, clearly seen in the increased density accreting opposite the largest plume.
The growth of the leading plume, and the strong accretion opposite it, give C15-3D a preferred axis \citep[see also][]{DoBuMu13}.

\section{Discussion}
\label{sec:discuss}

The role of dimensionality in CCSN modeling has been of intense interest recently starting with neutrino `lightbulb' models that parameterize heating with a fixed luminosity.
\citet{NoBuAl10} found that 3D \emph{enhanced} the potential for explosions, explaining the effect through increased dwell-time for parcels in the neutrino heating region.
\citet{HaMaMu12} were unable to confirm that result and for their base 3\degree\ simulations found little difference between 2D and 3D.
They found that improved angular resolution enhanced explosion in 2D and inhibited explosions in 3D, which they attributed to the differences in the turbulent cascade and the action of the SASI.
\citet{Couc13b} found that 3D \emph{diminished} the potential for explosions, while following the details of the first study carefully.
Subsequent studies with other parameterizations have mixed results.
Parameterizations necessarily omit physics that might be critical to the nature of the CCSN mechanism, thus the importance 2D versus 3D must be made relative to simulations containing all needed physics.
Most \nurhd\ simulations, including these, have shown that 3D diminishes the potential for explosion.

The earlier revival of the shock for C15-2D is consistent with the other \nurhd\ simulations in the literature.
Axisymmetric simulations of 11.2, 20-\msun\  \citep{TaHaMu13,TaHaJa14, Hank14} and 27-\msun\ progenitors \citep{HaMuWo13} produce explosions where their 3D counterparts do not.
\citet{MeJaBo15} obtained a delayed explosion in 3D by altering neutrino--nucleon scattering for the 20-\msun\ progenitor.
In the 3D 20- and 27-\msun\ models, they unambiguously demonstrate the spiral SASI mode identified by \citet{BlMe07}, whereas C15-3D resembles the neutrino-dominated simulations of \citet{AbOtRa14} and \citet{Fern15}.
In their multi-D simulations, the shock recedes, in the manner of 1D simulations.
Shock contraction in the \citet{HaMuWo13} 27-\msun\ model ends, when, after accreting a composition interface that drops the (shock) accretion rate from $\approx$0.75 to $\approx$0.25~\msuns, the shock rapidly expands to near its previous peak.
Following expansion, the shock revives in their 2D simulation, but contracts in their 3D simulation.
Likewise, in the 20-\msun\ simulation, a drop in accretion from $\approx$0.8~\msuns\ to $\approx$0.4~\msuns\ at 250~ms expandes \Rshock\ without explosion, and in the 11.2-\msun\ simulation a relatively flat accretion of $\approx$0.2~\msuns, results in a gradual decline of the 3D shock radius \citep{TaHaJa14}.
In \citet{MeJaBo15}, the same accretion drop in a 20-\msun\ 3D simulation triggers an explosion using enhanced neutrino heating. 
As in \citetalias{BrLeHi14}, sudden decreases in accretion at the shock, \accshock, did not trigger shock revival for this progenitor in 2D.
For our models at shock revival, \accshock\ is $\approx$0.7~\msuns\ for C15-2D ($\approx$250~ms) and $\approx$0.5~\msuns\ for C15-3D ($\approx$350~ms) and smoothly declining (Figure~\ref{fig:traces}d; dashed lines).

In the multidimensional models of \citet{TaKoSu14}, \Rshock\ rises smoothly beyond maximum 1D \Rshock, with the 2D simulations growing faster than 3D, though these models include less physics than the work of the Garching group discussed above and than our simulations.
In their highest resolution simulations all of the 2D runs and the longest-run 3D simulation show steepening of the \Rshock\ curve, as seen in our models.

In contrast to the above and to our results, \citet{MeJaMa15} find a modest \emph{increase} in explosion energy for 3D versus 2D for a low-mass 9.6-\msun\ Fe-core progenitor for which they also obtain an explosion in 1D.
They attribute this difference to reduced cooling due to increased dissipation of the accretion flow above the gain surface.
Given the progenitor differences (1D explosions, plumes not reaching the shock, etc.) these differences relative to our simulations are not surprising.

For the remaining 3D \nurhd\ simulations, including C15-3D, it appears that there is some sensitivity to neutrino heating and the pre-explosion shock history.
The 3D simulations of the Garching group \citep{HaMuWo13,TaHaMu13,TaHaJa14,MeJaBo15} all exhibit declining \Rshock\ that are only (temporarily) reversed by sharp declines in accretion.
Only by modifying the neutrino opacities to account for strange quarks does the $\sim$10--20\% increase in neutrino heating turn the shock reversal into an explosion.
Ostensibly, the Garching code is most similar to \chimera\ of all multi-dimensional \nurhd\ CCSN codes, though there are some differences:
we do not use the $\nue\nuebar \leftrightarrow \numt\numtbar$ pair-conversion process that increases cooling \citep{BuJaKe03};
we use the \citet{Coop85} nuclear EoS in the dynamically important shocked cavity;
they use variable Eddington tensor transport along rays while we use flux-limited diffusion;
and there are likely differences in numerical dissipation of flows and turbulence from angular and radial resolution and radial grid motion.
The effects of these differences can only resolved by code-to-code comparison.

\section{Summary and Conclusions}
\label{sec:summary}

We have computed two identically initialized CCSN simulations, one in axisymmetry (C15-2D) and the other without imposed symmetry (C15-3D), for 440~ms after core bounce using the \chimera\ CCSN code.
The shock revives fairly quickly ($\approx$250~ms after bounce) for C15-2D, following large $\ell=1$ oscillations of the shock from the SASI and buoyant plumes along the poles, similar to previous 2D simulations with \chimera\ \citepalias{BrMeHi13,BrLeHi14}.
Immediately preceding shock revival, C15-2D exhibits more neutrino heating in the gain region and higher accretion at the gain surface than C15-3D.
For C15-3D, the shock revival occurs $\sim$100~ms later, while accretion and convection continue.
Though these simulations do not extend long enough to determine final explosion energies, the energies reached in C15-3D, and their growth, make it likely that a lower final explosion energy will be reached in 3D, or it may take longer to reach a similar energy relative to 2D.

The shock is revived in C15-3D through the lead of a single, large-angle plume that results in expansion with a preferred axis, not wholly dissimilar to the axis-imposed structure of 2D simulations.
Based on examination of the buoyant plumes and their angular growth in C15-3D, we speculate that the development of fewer large-scale individual plumes may be necessary in 3D to provide the buoyancy needed to overcome the accretion ram pressure and relaunch the shock.

We see evidence for larger turbulent kinetic energy contributing to the earlier shock revival for C15-2D, but further examination is required.
Though these are among the best-resolved spectral-\nurhd\ CCSN simulations reported in the literature, we can not be certain the accretion streams and turbulence are adequately resolved.
We plan more extensive coverage of these issues, including an examination of resolution, in subsequent publications.

\acknowledgements
This research was supported by the U.S. Department of Energy Offices of Nuclear Physics and Advanced Scientific Computing Research;
 the NASA Astrophysics Theory Program (grant NNH11AQ72I); and the National Science Foundation PetaApps Program (grants  OCI-0749242, OCI-0749204, and OCI-0749248).
PM is supported by the National Science Foundation through its employee IR/D program. The opinions and conclusions expressed herein are those of the authors and do not represent the National Science Foundation.
This research was also supported by
an award of computer time provided by the Innovative and Novel Computational Impact on Theory and Experiment (INCITE) program at the Oak Ridge Leadership Computing Facility (OLCF) and 
at the Argonne Leadership Computing Facility, which are DOE Office of Science User Facilities supported  under contracts DE-AC05-00OR22725 and DE-AC02-06CH11357, respectively.
Animation by Mike Matheson at OLCF.

\end{document}